\documentclass[]{mn2e}
\usepackage{times}
\usepackage{lscape}

\begin{document}
\title[Multi-wavelength properties of Planetary Nebulae]
{Multi-wavelength diagnostic properties of Galactic Planetary Nebulae detected by GLIMPSE-I}

\author [M. Cohen et al. ]
{Martin Cohen$^{1}$, Quentin A. Parker$^{2,3}$, Anne J. Green$^{4}$, Brent Miszalski$^{2,5}$, David Frew$^{2}$,
\newauthor Tara Murphy$^{4,6}$\\
$^{1}$Radio Astronomy Laboratory, University of California, Berkeley, CA 94720, USA\\
$^{2}$Department of Physics, Macquarie University, Sydney, NSW 2109 Australia\\
$^{3}$Anglo-Australian Observatory, PO Box 296, Epping, NSW 2121, Australia\\
$^{4}$Sydney Institute for Astronomy, School of Physics, The University of Sydney,
  NSW 2006, Australia\\
$^{5}$Centre for Astrophysics Research, STRI, University of Hertfordshire, College 
   Lane Campus, Hatfield  AL 10 9AB, UK \\
$^{6}$School of Information Technologies, The University of Sydney, NSW 2006, Australia\\}

\date{ Accepted . Received ; in original form        }
\renewcommand{\baselinestretch}{1.6}

\maketitle

\pubyear{2010}
       
\begin{abstract}
We uniformly analyze 136 optically detected PNe and candidates
from the GLIMPSE-I survey in order to to develop robust, multi-wavelength, 
classification criteria to augment existing diagnostics and provide 
pure PN samples. PNe represent powerful astrophysical probes. They are important dynamical 
tracers, key sources of ISM chemical enrichment, windows into 
late stellar evolution, and potent cosmological yardsticks. But 
their utility depends on separating them unequivocally from the many
nebular mimics which can strongly resemble {\it bona fide} PNe in
traditional optical images and spectra. We merge new PNe from
the carefully evaluated, homogeneous MASH-I and MASH-II surveys,
which offer a wider evolutionary range of PNe than hitherto available, 
with previously known PNe classified by SIMBAD.
Mid-infrared (MIR) measurements vitally 
complement optical data because they reveal other physical processes and 
morphologies via fine-structure lines, molecular bands and dust.  MIR
colour-colour planes, optical emission line ratios and radio fluxes show 
the unambiguous classification of PNe to be complex, requiring 
all available evidence.  Statistical trends provide predictive 
value and we offer quantitative MIR criteria  to determine 
whether an emission nebula is most likely to be a PN or one of the 
frequent contaminants such as compact H{\sc ii} regions or  symbiotic systems. 
Prerequisites have been optical images and spectra but MIR morphology, colours, environment 
and a candidate's MIR/radio flux ratio provide a more rigorous classification.  
Our ultimate goal is to recognize  PNe using only MIR and radio characteristics,
enabling us to trawl for PNe effectively even in heavily obscured regions of the Galaxy.
\end{abstract}

\section{Introduction}

Planetary nebulae (PNe) are the resolved, ionized gas shrouds
of dying stars. PNe are expected to be numerous as their progenitors are
the evolved cores of stars once
in the mass range 1-8 times that of the sun that represent 90\,percent of all
stars more massive than the sun.  Estimates suggest  that there may be
as few as 5,000 (Moe \& De Marco 2006a) but as many as 24,000 PNe (Frew, 2008)
in our Galaxy although fewer than 3000 are currently known
(Frew \& Parker 2010). The role of PNe in our Galaxy is significant since 
its chemical evolution depends critically on the amount of enriched 
mass recycled by PN progenitor stars
via all mass loss episodes including the final PN ejection. These ejecta
are significantly enriched by stellar nuclear processes. In the  first 
giga-year of galaxy evolution supernova explosions and  high-mass PN progenitors 
return about the same gross mass but thereafter the PN contribution 
increasingly dominates. Low and intermediate mass stars are an additional  
source of pre-biotic carbon  later in the interstellar medium (ISM), returned in their PN phase, 
although massive stars return more (Gustafsson et al.  (1999, Henry, Edmunds \& K\"oppen 2000).
However, the thin disc and metal-rich thick disc of our Galaxy gain most
carbon from low- and intermediate-mass stars (Bensby \& Feltzing 2006)
as do any regions with higher metallicity (Chiappini et al.  2003).

PNe exhibit strong emission lines making them detectable to
great distances and this allows determination of accurate velocities and
expansion rates (from which PN dynamical ages can be gauged) across the
entire Galaxy.   Importantly, the luminosity distribution of PNe in 
external galaxies allows their use as distance calibrators
in determining the scale of the Universe to an accuracy of about 10\,percent (e.g.
Jacoby 1989, Ciardullo et al. 2002, 2006; Ciardullo 2010). This luminosity
distribution encodes rich information on our Galaxy's star-forming history.
More parochially, the complex morphologies of PNe provide clues to their
formation, evolution, mass loss processes, and the shaping role played by binary
central stars or even sub-stellar objects and planets (Soker (2004), Miszalski 
et al.  2009a). Also, as the central star rapidly fades to become a white 
dwarf and the nebula expands, the integrated flux, surface brightness and 
radius all change in a way that can be predicted by current stellar and 
hydrodynamic theory.

In all these ways, PNe are effective astrophysical probes which have
particular utility if their distances are also known (currently a major
problem in the field). We have made recent progress
here with the adoption of a new surface-brightness radius relation that
shows great promise (e.g Frew 2008, Frew \& Parker 2010), as we
demonstrate later.

However, a major problem that has undermined the value of previous PNe
catalogues is the presence of contaminants.
Precise PN identification is complicated by the wide variety of
morphologies, ionization characteristics and surface
brightness distributions of the PN family which reflects the stages of
nebular evolution, progenitor mass and chemistry and the possible
influence of common envelope binaries (e.g. De Marco 2009, Miszalski 
et al.  2008a, 2009a).  Frew (2008) for example, has recently shown 
that up to 15\,percent of objects currently accepted as nearby PNe (e.g.
Hewett et al.  2003) are actually just ionized regions of the general
ISM around hot white dwarf or subdwarf stars 
(e.g. Frew \& Parker 2006, Frew \& Parker 2010).

Our group has now tested and adopted a range of criteria to 
eliminate such contaminants more effectively.
Only the recent on-line availability of imaging surveys and other data has
enabled such clear discrimination tools to be developed (e.g. Parker et
al. 2006; Madsen et al. (2006), Frew \& Parker 2010).
In this paper we particularly emphasize the power of the newly available
high quality, high-resolution and deep MIR imaging  in
the Spitzer GLIMPSE survey  to contribute to robust PN identification.

The field of MIR work on PNe is highly active now, with contributions from  
Chu et al.(2009); Hora et al. (2004),  Kwok et al. (2008); Madsen et al. (2006), 
Phillips \& Ramos-Larios 2008,2009), Phillips \& Zepeda-Garcia (2009), and
Ramos-Larios \& Phillips (2008).
Cohen et al.  (2007a: hereafter Paper-I) undertook a preliminary initial
analysis of MASH-I PNe (Macquarie-AAO-Strasbourg H$\alpha$ PN Project:
Parker et al. 2006),  observed by the Spitzer GLIMPSE project. Optical, MIR, and radio 
characteristics were examined and combined. Although the results were limited 
to a sample of only 54 PNe, Paper-I had examples of objects rejected as non-PNe 
and adduced reasons for their rejections.It probed the detailed spatial 
relationships between H$\alpha$ and MIR images, and it established an 
ensemble value for the MIR/radio flux ratio in PNe that was almosts the 
same as we derive here.  The current paper follows on from Paper-I.
In the intervening period, the MASH-II catalogue has been created by
Miszalski et al.  (2008b). Unlike MASH-I PNe, which were discovered on
the AAO/UKST/ H$\alpha$ survey by visual methods,
the many compact MASH-II nebulae were found by semi-automated
techniques applied to the image parameterization data from the digitized
version of the original photographic film material, e.g. Hambly et al. 
(2001). Seventy-five percent of new  MASH-II PNe were  found using the
Image Analysis Mode of the SuperCosmos data pipeline (Hambly et al. 
(2001) to target star-like or compact PNe. The remaining 25\, percent 
are highly evolved, very low surface brightness, extended PNe, 
uncovered only by quotient imaging of
the factor 16 blocked down matching H$\alpha$ and R-band exposures. 
A fully detailed description of MASH-II discovery techniques
is given by  Miszalski et al.  (2008b).  The verification of nebulae as  {\it bona fide}
PNe is primarily achieved by  follow-up ground-based optical spectroscopy, exactly
as for MASH-I. Hundreds of MASH-II PNe candidates were discovered
and currently 360 are classified as true (T), likely (L) or
possible (P) PNe following a similar procedure to that for MASH-I.  The
combined MASH-I and MASH-II catalogues cover the same region of
the Galaxy and together effectively span the full evolutionary range of
PNe from the young, high density compact PNe  to those dissipating 
into the ambient ISM.  Although MASH-I and MASH-II probe the same 
Galactic volume and include the same 
types of nebulae, the respective fractions of compact PNe and 
extremely evolved PNe are different. An objective of the current 
paper is to compare the multi-wavelength attributes of
the two sets of MASH PNe and previously known PNe.  We also emphasize the vital importance of
a rigorous uncontaminated sample of PNe. This is done by utilizing the
discriminatory power inherent in homogeneous multi-wavelength 
surveys, and by developing associated diagnostic criteria with 
strong predictive capability. Some cautionary notes on traps
that occur in applying sweeping generalizations with limited 
data are also made. 

In \S2 we describe our selection of nebulae and discuss techniques to
remove non-PNe
from existing PNe catalogues. \S3 presents our sample of PNe which
{we believe} constitutes a ``clean" sample of 136 nebulae
{to satisfy our stringent selection criteria.
\S4 discusses the spatial distributions of MIR emission in nebulae of
large apparent diameter. \S5 addresses the classification of 
nebular morphologies. \S6 offers a series of analyses of PNe based 
on measured IRAC MIR colours, optical emission-line flux ratios,
IRAC false colours, radio continuum fluxes and MIR/radio ratios.   In
\S6 we also compare the IRAC colours of PNe,
drawn from several subgroups, with the colours of  diffuse and compact and
ultra-compact H{\sc ii} regions (UCHIIs)  and we explore IRAC 
colour-colour planes and determine which combination of
colours might best
separate PNe from contamination by other kinds of source. Another objective
is to seek the largest metric
distances in the colour-colour planes between the locations of PNe and
of their contaminants, and we demonstrate statistically significant quantitative
discriminants between PNe and objects frequently confused with PNe.
\S7 investigates whether the diagnostic diagram of common emission line
ratios first proposed by Sabbadin, Minelo, \& Biancini (1977: SMB)
has any utility for this set
of PNe when separating them by their discovery method, optical spectral lines, their MIR
false colours, age, or their morphology. \S8 examines the properties of our PNe in relation
to their derived ages. \S9 examines the radio properties and  the median MIR/radio flux ratios
for different subsets of PNe and in relation to age. \S10 seeks the 
 possibility of a link between MIR false colours and
nebular excitation, PN type and MIR/radio ratio.
\S11 compares the colours of Galactic and LMC PNe. 
\S12  presents our conclusions.

The appendix  summarizes those objects which we, and others, have rejected as
PNe, in the hope that this list will help PN investigators to drop 
these contaminants from their own studies of PNe. These objects should 
be withdrawn from the PN literature, principally from SIMBAD, although 
some of this updating has already occurred with the incorporation of 
MASH nebulae into Vizier.  

\section{Sample selection}
The defining criteria of our sample are i) that all objects have been
assigned to the category of PNe
either in the new  MASH-I or MASH-II databases (available through Vizier)
or via current listing in SIMBAD as a PN and ii) must lie within the area
covered by GLIMPSE-I  (Benjamin et al.  2003; Churchwell et al.  2009), and iii)
within the Galactic longitude boundaries of
the SuperCosmos H$\alpha$ Survey of the Southern Galactic plane (SHS: Parker
et al.  2005). The SHS was the discovery medium for MASH and, therefore, all
these PNe have optical counterparts. This is unlike the possible PN
identified at (313.355, +0.312), reported by Cohen et al.  (2005)
based solely on radio and MIR data since no meaningful optical emission
could be established due to excessive extinction along the PN sight line. 
\footnote{We have recently noted possible very faint H$\alpha$ emission 
in a quotient image of the vicinity of the IR ring.  This is too weak to 
define any morphology and no optical spectrum is yet available.} 
Finally, together with the GLIMPSE longitude limits of 10$^\circ$ 
to 65$^\circ$, and 295$^\circ$ to 350$^\circ$}, we also chose to cut 
at Galactic latitudes of $\pm$1.16$^\circ$. The
boundaries of the GLIMPSE area are jagged beyond $\pm$1.0$^\circ$ in
latitude because the individual Spitzer IRAC frames were not laid down in the
sky parallel to Galactic coordinate axes. Twenty-one PNe 
close to the latitude limits of GLIMPSE had incomplete wavelength coverage
(3.6 and 5.8\,$\mu$m, or 4.5 and 8.0\,$\mu$m).
In Paper-I the latitude range of the PN sample was $\pm$1.0$^\circ$.  
Consequently our  sample of MASH-I PNe with the new, broader limits was 
enlarged to  66 objects; 12 more than for Paper-I. It also
includes 31 new MASH-II nebulae. Note that a number
of PNe initially discovered by MASH  have subsequently entered
the literature under other designations while MASH databases were being
finalized.

Our third sample consists of 44 PNe recorded in the literature prior to the
MASH surveys. This sample represents the heterogeneous ``known'' PNe compiled historically
from disparate sources into, for example, the Acker (1992, 1996) and
Kohoutek (2001) PN catalogues.  These were added as these previously known
putative PNe may not be representative of the full PN evolutionary
range compared to MASH  nebulae and also because they have not been filtered
through the same diagnostic criteria used to identify PNe in MASH and,
therefore, may contain higher levels of contaminants.  After Acker's catalogues and 
during the observing period of MASH, many papers appeared containing small 
numbers of new PNe and some of these non-MASH PNe look very much 
like those preferentially discovered by MASH; for example, the objects found 
by  Beer \& Vaughan (1999), Cappellaro et al.  (2001), and Boumis et al.  (2003, 2006).

\subsection{Removing contaminants from PN catalogues}
\subsubsection{Identifying MASH contaminants}
Although careful assessment of candidate PNe was made in compiling MASH
catalogues in the first place (Parker et al.  2006, Miszalski et al.  2008b) they
have been critically re-examined here based on a combined
multi-wavelength approach incorporating GLIMPSE data,
rather than via assessment of merely optical and NIR/MSX attributes (e.g.
Cohen et al.  2007a).  Indeed,
as higher quality, deeper, optical spectra are acquired by the MASH
consortium for poorly studied MASH PNe,
lines too weak to have been recognized in the original
confirmatory spectra of the nebulae can call into doubt the original PN
classification, especially for compact or barely resolved objects. As a
result several  MASH objects have been re-classified as symbiotic
stars, e.g. PHR1253-6350 (Paper-I).

One technique used for original MASH culling was based  on 2MASS colour-colour plots, e.g. Schmeja \&
Kimeswenger (2001),   Corradi et al.  (2008).  Coding 2MASS  $J,H,K_s$ images
as
blue, green and red, respectively, then combining them into a   false-colour
composite can also be a valuable discriminator. True unresolved PNe appear
violet in 2MASS false colours while  resolved PNe are often purple in 2MASS.
In general this is due to the inclusion of the strongest  emission  lines in the J-band
(1.25\,$\mu$m) and  K$_s$ band (2.16\,$\mu$m).  The relevant expected PN
lines were listed by Whitelock (1985) who forecast the great strength of the He{\sc i}
triplet at 1.083$\mu$m. This prediction has been well demonstrated  (e.g. for
NGC\,7027, by Treffers et al.  1976). The IR spectral line content of PNe was
revisited
by Hora et al.  (1999) on the basis of 1$-$2.5-$\mu$m spectroscopy.  These
authors divided their PN sample into groups, according to the dominance 
of the continuum or emission lines of H{\sc i} and/or H$_2$.

These characteristics determine the strongest
emission lines in the $JHK$ bands. For example, in H{\sc i} line-dominated
PNe, for $J$, Paschen$\beta$ is the brightest line, augmented by 
1.083-$\mu$m He{\sc i}, 1.26-$\mu$m [Fe{\sc ii}], and1.316-$\mu$m 
O{\sc i}; for $H$, the H{\sc i}  Brackett series dominates, with
contributions from 1.700-$\mu$m  H{\sc i} and 1.644-$\mu$m 
[Fe{\sc ii}] (the latter
in PNe with internal shocks); for $K$, Br$\gamma$, He{\sc i} at 2.058 and
2.112-$\mu$m, with bright 2.122-$\mu$m of H$_2$ present in some PNe (see \S4).

Another major source of PN contaminants are  diffuse, compact, and UCHIIs.
If high signal-to-noise optical spectra are lacking it may prove
impossible  to discriminate
between some very low excitation PNe and some H{\sc ii} regions because the
relative strengths of the [O{\sc iii}], H$\beta$, H$\alpha$, and [N{\sc ii}]
lines exhibited by both types of
nebula can overlap (e.g. see Frew \& Parker 2010).  Our multi-spectral
approach to nebular discrimination and the various spectral and photometric
signatures for most PNe contaminants has recently been reviewed by Frew \&
Parker (2010).

In this paper we specifically reveal how the MIR morphology of nebulae can
provide a valuable additional discrimination tool because true UCHIIs
are only rarely unresolved by IRAC at 8\,$\mu$m (Cohen et al.  2007b).

\subsubsection{The discriminatory power of MIR and radio data}
The presence of filaments, extended structures and/or an amorphous
halo in the MIR imagery, exterior to the optical image, generally indicates 
an H{\sc ii} region rather than a PN. We draw a clear distinction between 
the generally symmetrical, roundish photodissociation regions (PDRs) of PNe and the irregular 
streamers that Cohen et al. (2007b identified.  An excellent example of 
such a streamer is given by these authors in their Fig.18. 
False color also plays a role.  PNe are found only in three false colours, 
none of them matched by the false colour of any HII region in the MIR.

Compact PNe which are isolated in the MIR are easy to identify but not every
MIR-detected PN is compact nor isolated from other more extended MIR
emission.  Creating MIR false-colours images
by representing the three IRAC bands at 4.5, 5.8, and 8.0\,$\mu$m by
blue, green, red, respectively (hereafter IRAC false colour or MIR false
colour) is also a valuable discriminatory
tool (Paper-I) to identify PNe in more complex regions.  For example, in
their blind 20-GHz radio survey of the  Galactic plane for UCHIIs,
Murphy et al.  (2010) noted that about 10\,percent of their sample were easily
recognized as PNe  on the basis of their IRAC false
colours which appear red, as opposed to the yellow (PAH emission) or
white (broadband thermal emission by dust) of H{\sc ii} regions. This
discrimination is further assisted via other indicators such as the
lack of obvious association of true PNe with amorphous extended MIR
structure and their low observed ratios of MIR to radio continuum  
(e.g. typically a value of $\sim5$ compared with the median ratio of 42 for UCHIIs).
The formal difference between these two independent medians is significant at the
5\,$\sigma$ level (where $\sigma$ combines the errors of these median ratios in quadrature). 
For PNe with well-known distances (e.g. such as those  in the
Magellanic Clouds), a large radio flux density ($\geq$\,5\,mJy) is found
to be a promising discriminant between H{\sc ii} regions and PNe  which
typically are weak radio emitters (e.g. Filipovic et al.  2009).  
In short, there is an increasing number of criteria that
can be used to characterize a nebula as a PN or contaminant in addition to
the classical reliance on its optical morphology and spectrum.  In their
recent discussion of the changing landscape of PN classification Frew \&
Parker (2010)  emphasize a problem with investigating extragalactic PN
candidates, namely that high-resolution optical imagery is often lacking,
rendering MIR and radio criteria even more important as parallel indicators.

Young stellar objects (YSOs) have also been cited as confused with known PNe
(Whitney et al.  2008) in an LMC colour-magnitude diagram. The grid of precomputed
energy distributions of YSOs (Robitaille et al.  2006) contains 200,000
spectra, and spans too vast an area to represent even by median colours.  However,
 in order for a YSO spectrum to mimic that of a PN requires a large
extinction, too large for any object discovered optically, and much more plausible
for a YSO embedded in a dark cloud. Visual examination of the environs is usually
adequate to distinguish a potential YSO from a PN.

Honing this new set of multi-wavelength discrimination tools so that one
gains experience of the relative weighting of these diverse criteria is an
implicit goal of the present paper. We will emphasize the potential role
of MIR properties of nebulae and their locations in colour-colour diagrams 
as further means both to eliminate contaminants from PN surveys and to refine
traits and evolutionary characteristics of  {\it bona fide} PNe.

\subsection{Examples of ambiguous objects culled from the sample}
There are many nebulous objects in the extant PN catalogues which have been
notoriously   difficult to classify by any number of previous litmus tests
because  the existing observational evidence is inconclusive or even
contradictory. They often have conflicting identifications in SIMBAD.
We select two such objects here to illustrate the additional discriminatory
capabilities of the GLIMPSE MIR data that can finally shed light 
on the true nature of such objects.

\subsubsection{Hen2-77}
\begin{figure}
\vspace{7cm} 
\includegraphics{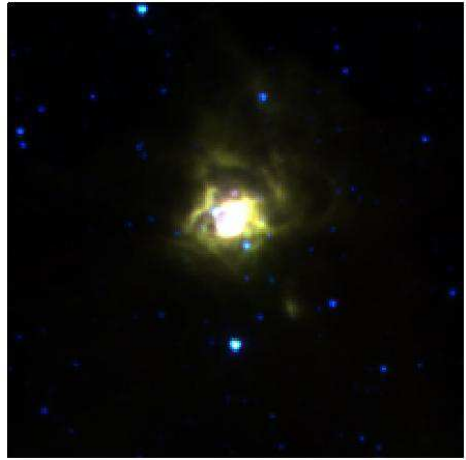}
\caption{IRAC False colour image of Hen2-77 reveals the MIR morphology to be that of an UCHII with 
surrounding filamentary structure, and not a PN. Image is 4$^\prime$ $\times4^\prime$. Galactic
N and E are up and left.\label{hen234}}
\end{figure}

\begin{figure}
\vspace{6cm}  \includegraphics{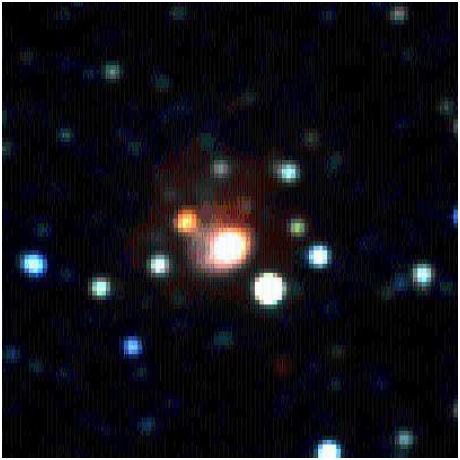}
\caption{2MASS false colour image of  Hen2-77. Image is 100$^{\prime\prime}$ $\times100$ $^{\prime\prime}$. 
Galactic N and E are up and left. \label{henjhk}}
\end{figure}

The first such object is Hen2-77 which has a long history of controversy
in the literature. Roughly half the 75 articles in SIMBAD in which
it is mentioned regard it as a PN.  However, Cohen \& Barlow (1980)
concluded that it is a heavily reddened H{\sc ii} region because of its
patchy, irregular optical appearance, lack of an  identified
central star, large radio continuum flux density, and
its  high A$_V$ of 9, derived from the Balmer decrement.  IRAC false colour
imagery (Fig.~\ref{hen234}) further reinforces
the argument in favour of an H{\sc ii} region nature due to the
character of its MIR morphology  (Cohen et al.  2007b).  Its 2MASS false
colour counterpart (Fig.~\ref{henjhk}) shows an orange
and white diffusely extended source rather than the normal purple colour of
many PNe caused by the combination of strong [Fe{\sc ii}] in the J-band 
and the S1\,1-0 H$_2$ line in the $K_s$ band.

Hen2-77 is indistinguishable from the many other sources in the surveys
of H{\sc ii} regions by Caswell \& Haynes (1987) and by Kuchar \& Clark
(1997), and other references to it as a thermal radio source can
be found by examining the remainder of the 75 citations in the
literature that do not conclude that it is a PN, particularly the more recent
references since 2000, where its MIR spectral character is discussed.
For example, Peeters et al.  (2002) include it under its IRAS name, IRAS\,1206$-$6259, 
(it is also known as RAFGL 4144) in their MIR spectral catalogue
of 45 compact H{\sc ii} regions. These authors derived a luminosity of
4.5$\times$ 10$^{5}$ L$_{\sun}$, clearly marking it as an H{\sc ii} region.

\begin{figure}
\vspace{7cm}
\includegraphics{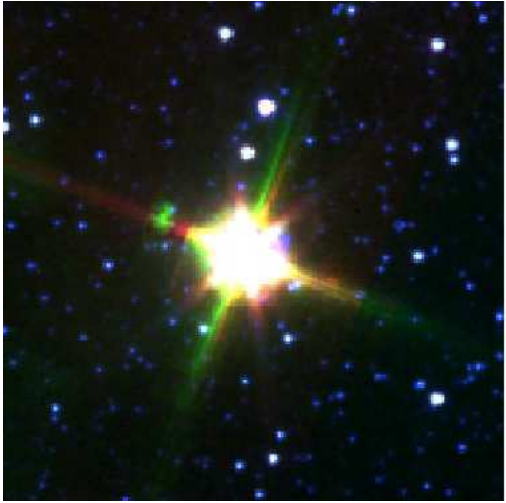} 
\caption{Mz~3 in IRAC false colour. The Spitzer diffraction spikes indicate the
presence of a very compact, but unsaturated, central core. Image is 4$^\prime$ $\times4$ $^\prime$. Galactic
N and E are up and left. \label{Mz3}}
\end{figure}

\begin{figure}
\vspace{20cm}
\includegraphics{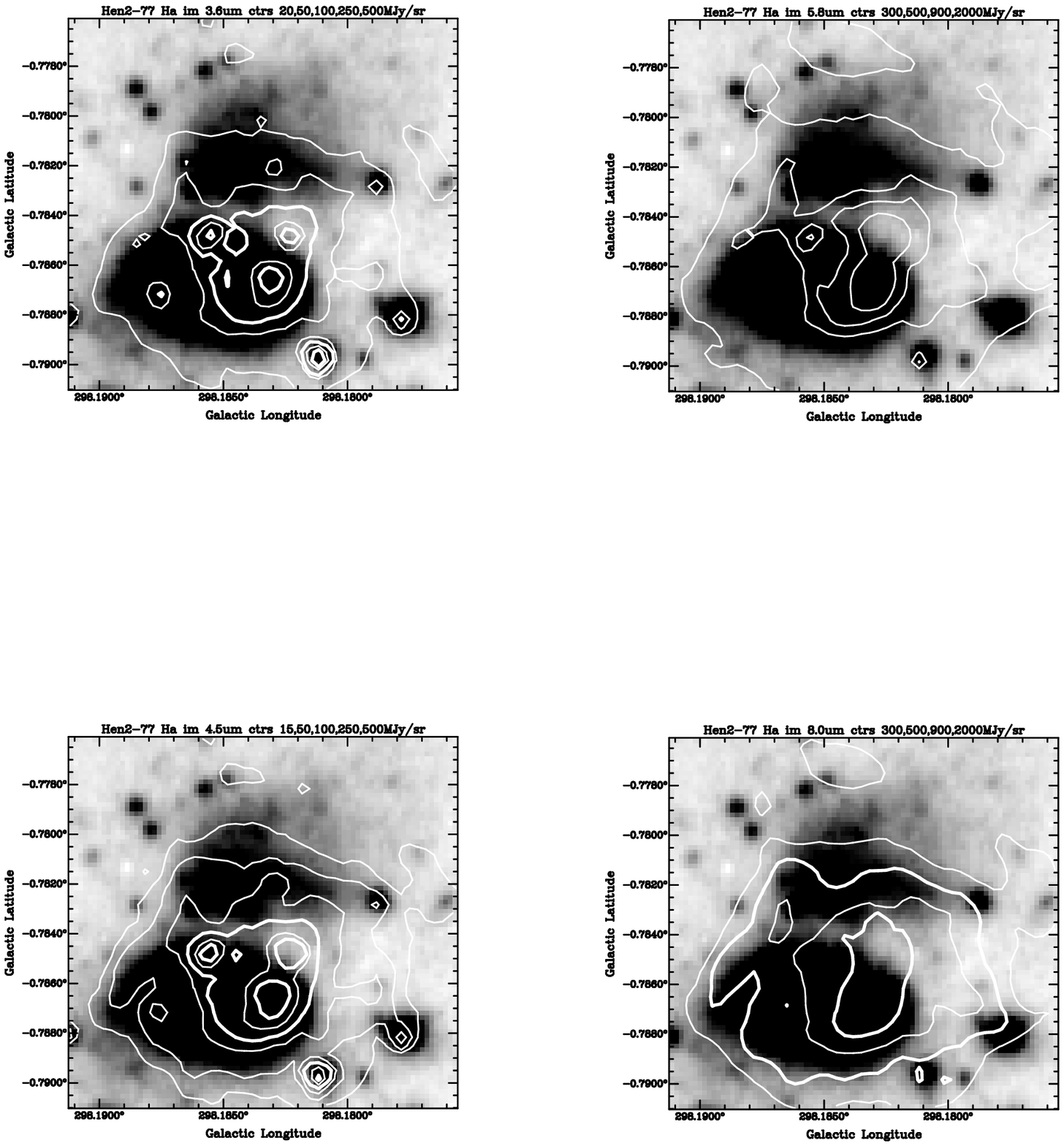}
\caption{``Quartet'' of IRAC MIR images in white contours overlaid on the
grey scale H$\alpha$  image of Hen2-77.  IRAC bands at 3.6, 4.5, 5.8, 8.0\,$\mu$m
are overlaid in the top left; bottom left; top right; and bottom right
positions. Note the dark lane (i.e. pale grey in the H$\alpha$ images) that
divides the H$\alpha$ nebulosity into two separate pieces. For each IRAC band
the white contour levels in MJy/sr are given at the top of the figures.
\label{he277qtet} }
\end{figure}

There appear to be two separate nebulosities in the optical yet only
a single MIR extended object that embraces both these
nebulae.  A dust lane might account for this.The dominant MIR core lies
to the south-west and it contains a number of seemingly embedded point
sources, perhaps a nascent cluster of
stars.  Unlike many PNe, the four IRAC images of Hen2-77 (Fig.~\ref{he277qtet})
have the same size and structure.  The IRAC false colour image
(Fig.~\ref{hen234}) shows a central  white rhomboid,
implying emission in all three bands and indicative of continuum thermal
emission by dust grains.  The core is enveloped by several faint, yellowish
(probably PAH-emitting) filaments, very similar to other compact H{\sc ii} 
regions (Cohen et al.  2007b) and UCHIIs (Murphy et al.  2010) 
but atypical of PNe.  The
yellow-green colour suggests PAH emission in the IRAC 5.8 and 8.0-$\mu$m
bands, consistent with its MIR
spectrum (e.g. Cohen et al.  1989), but clearly different from the orange
and/or red false colours of PNe that are
attributed to PAHs or H$_2$ lines.  The ratio of 6.2-$\mu$m to 7.7-$\mu$m
intensities of the PAH features is 40\,percent lower for PNe than for H{\sc ii}
regions (Cohen et al.  1989, their Table 3) and this distinguishes the red
false colour of PNe from the yellow of H{\sc ii} regions with PAHs.

Finally, the MIR/radio ratio is 42, again typical of UCHIIs
 and much larger than the ratio for any group
of PNe (e.g. Table~\ref{mirrad}, \S9.1, and Paper-I).
This, too, suggests that the object is, indeed a  compact H{\sc ii} region
and that the ambiguous nature of this object has finally been resolved.

Given that there are so many arguments against Hen~2-77 being a PN,
one might wonder how there could have been so much confusion as to
its character. The first description of its optical spectrum was by
Henize (1954, 1967).  Weak H$\alpha$ was seen and the line was
``definitely widened relative to other nearby emission lines
(`diffuseness' = 2)''. No continuum was detected even in a long
exposure. These two criteria, applied to 95 PNe deemed to be
well-confirmed, constituted the basis for its PN classification.
Confirmation was further based on a resolved, reasonably regular,
nebular image, however, no blue lines were recorded, probably due to
the strong reddening.   Henize stated the purity of his PN sample
to be 97 per cent, but Westerlund \& Henize (1967) described it as a
``peculiar object'' with a large bright knot on the west. In short,
the classification rests solely on a weak H$\alpha$ emission line and
the absent continuum, but the irregular image should have raised
suspicion about the certainty of a PN classification.

The spectroscopic detection of moderately strong [OIII] emission was
noted by Acker et al. (1992) and Kingsburgh \& Barlow (1994)
([OIII]/H$\beta \simeq$ 7), which leads to the conclusion that this
is a high-excitation compact HII region.  The apparent presence of
HeII $\lambda$4686 emission noted by Shaw \& Kaler (1989) suggests it
may hold one or more embedded Wolf-Rayet stars.

\subsubsection{Mz~3 ``The Ant" nebula}
The second ambiguous object is Mz~3, whose literature,
by  contrast, currently consists of 265 references in SIMBAD, almost 80\,percent
of which treat this unusual, highly collimated bipolar nebula as a
planetary.  The central core is unsaturated in IRAC and white in false 
colour, suggesting that warm dust emission is present, as found in many 
H{\sc ii} regions.  However, this is also a characteristic
of bipolar PNe that arises from the strong temperature gradients within
their equatorial dust disks. The large dusty disk around Mz~3 was 
first identified by Cohen et al.  (1978).

Fig.~\ref{Mz3} shows Mz~3 in IRAC false colour, with the dense dust disk
around the core in white (thermal emission by heated dust).  The Spitzer 
diffraction spikes indicate the presence of a very compact, but unsaturated, 
central core.

Outside the bright core the putative PN is red in  IRAC false colour.  The
polypolar PN NGC\,6302 is in many ways similar to Mz~3, and
so, by association, Mz~3 would also appear to merit the PN
designation. It too has a large dust disk (Lester \& Dinerstein 1984) and in MIR
false  colour its core is white and the surroundings are orange. The
MIR/radio ratios are 23 for Mz~3 and 16 for
NGC\,6302. These values are well above the median for PNe (this paper) but
very close to the median value (25$\pm$5) for  compact and diffuse H{\sc ii} regions (Cohen et
al. 2007b). Another important aspect might also militate against assigning
Mz~3 to PN status. 

\onecolumn

\twocolumn
\noindent
The 2MASS colours of the bright core of Mz~3 are
$J-H$=2.00, $H-K_s$=1.75, locating it among the dusty symbiotic stars in the
colour-colour plot of Corradi et al.  (2008), although a few ``true"
PNe also share these colours. In \S7 we shall see that Mz~3 is also highly 
anomalous in its optical emission line ratios. The possible analogue NGC\,6302 
is too extended to have reliable 2MASS photometry so it cannot be placed in 
the same NIR colour-colour plane.   It is instructive to compare the spatially 
integrated absolute K-magnitudes of the most luminous known PNe
with that of Mz~3, which has M$_K$ in the range $-$6 to $-$7, depending on the
adopted distance (e.g. 1.6$-$2.5 kpc from the literature). By contrast,
typical luminous PNe have M$_K$ between $-$3 and $-$4 (e.g.Wainscoat et al. 
1992, their Table\,2).   The bright, solar neighbourhood PN, NGC\,7027, has
M$_K$ of $-$3.7; NGC\,6302 $-$3.0; and even IR-[WC] PNe such as  He\,3-1333 have
only M$_K$ $\simeq$ $-$4. The large, additional 2.2~$\mu$m luminosity in Mz~3 
has been attributed to a cool AGB companion star (Smith 2003). Chesneau et al. 
(2007) find no evidence of a Mira companion, as suggested bySchmeja \& Kimeswenger (2001),
and prefer to associate Mz~3 with a young version of binary post-AGB stars.
Mz~3 has also been considered as a link between D-type (dusty)
symbiotic stars and PNe.
For example Lutz et al.  (1989) regard Mz~3, Hen2-104 and NGC 6302 as
representing a small group of bipolar PNe which they argue   are in
transition from a PN to a symbiote.  These authors identify low ionization
objects like Mz~3 and M2-9 as having a dust torus still sufficiently opaque
to prevent ionization of the outer disk, and in a phase preceding the 
high ionization PNe like NGC\,6302. Santander-Garcia et al.  (2008)  find 
an unexpectedly large  ionized gas mass of 0.1M$_\odot$ for Hen2-104.  
Yet  Smith \& Gehrz (2005) estimate the total mass of gas in Mz~3 
as 0.6M$_\odot$. Consequently, Mz~3 is unusual even among 
those objects with which it has the closest affinity.

Frew \& Parker (2010)
relate objects like M~2-9 and Mz~3 to the symbiotic B[e] stars. While we cannot yet
resolve the true nature of Mz~3, due to its non-conformist multi-wavelength
characteristics, it seems prudent to exclude it from our previously known PN
sample, especially as its MIR/radio ratio places it squarely in the
regime of H{\sc ii} regions.  

\begin{figure}
\vspace{9cm}    
\includegraphics{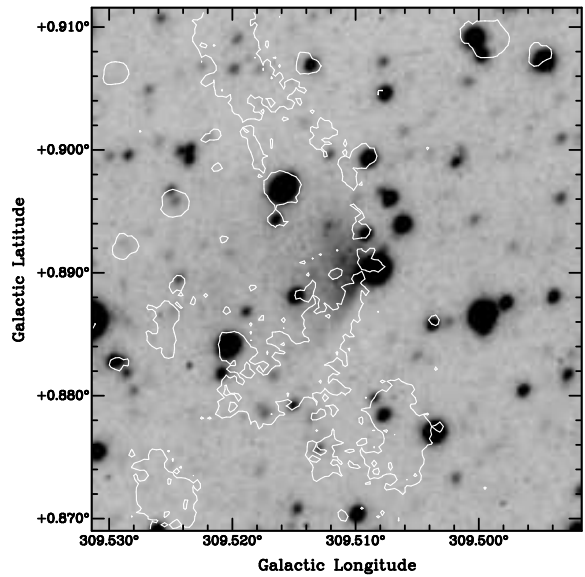}
\caption{Rejected PN candidate PHR1346-6116: H$\alpha$ greyscale; white contours of 8.0\,$\mu$m
emission.  The nebula is just visible as a faint diffuse darkening at the 
centre of the image, slightly east of the inflection in the diffuse 
8.0-$\mu$m arc.  \label{bowshock}}
\end{figure}

\subsection{Other culled nebulae}
The unique object G313.3+00.3  (Cohen et al.  2005) is culled because only extremely  faint optical data are
available for this highly obscured nebula that was discovered purely through
radio and MIR observations, despite the fact that it is highly likely to be
a {\it bona fide} PN (see footnote in \S2).
We have also eliminated several objects that have the appearance of bow
shocks. Two of these are MASH candidate PHR1346-6116, and the previously known
object PN G309.5+00.8. Such bow shocks might be associated with cataclysmic variables 
or simply represent random density
enhancements in the diffuse ISM (Frew \& Parker 2010: sec.4.10) that are 
ionized by nearby, reddened, hot,  field stars
that are passing through (Fig.~\ref{bowshock}). A large 8.0-$\mu$m
arcuate structure in the ISM is indicated by the curving white contours from
north to south in this field.  This arc abruptly changes direction at   the
location of the H$\alpha$ nebulosity.

Without the often definitive and corroborating evidence of an optical
spectrum no candidate can  yet be treated as a  {\it bona fide} PN. 
Consequently, we have not included any of the candidates offered by 
Phillips \& Ramos-Larios (2008) that fall within our survey bounds
although one or two, based solely on optical morphology, appear to be good
PN candidates. It is possible that one might 
eventually be able to substitute NIR and/or MIR spectroscopy
for definitive optical spectral confirmation when obtaining the latter is
problematic, due to a high level of extinction.

Note that many hundreds of IRAS sources have been suggested as potential
PNe solely on the basis of their far-IR colours and they often lack any
optical counterparts.  However, the limited follow-up efforts to date 
have provided a very low yield  of  {\it bona fide} PNe.
Su\'arez et al.  (2006) pursued 253 IRAS source PN candidates but were able to
confirm only 36 as PNe on the basis of optical spectroscopy.   Ramos-Larios
et al.  (2009) have returned to IRAS sources but add NIR data to seek 
heavily reddened candidate PNe.  No confirmed PNe are reported yet.
Kwok et al.  (2008) have also examined GLIMPSE images and presented a list of
30 objects which they assign to the status of PNe.  Sixteen objects are
indeed confirmed as such; 6 are in MASH and 10 are in the earlier
literature. However, we believe that the remaining 14 nebulae are not in
fact PNe. We are able to make this identification based on use of our new
and existing discriminatory tools.  Firstly, half of the objects we reject
as PNe lack the critical verification through optical spectroscopy though we
note seven had  already been rejected by Parker et al.  (2006) on the grounds
of their morphology and optical spectra. Three of the sample were also
explicitly  identified as H{\sc ii} regions by Churchwell et al.  (2006) in
their catalogue of 322 thermal bubbles found in GLIMPSE.  Another key point
is that seven of the 14 objects we have rejected have very large integrated MIR flux
densities at 8.0\,$\mu$m ranging from 66 to 713\,Jy which fall far outside
the range found for confirmed Galactic PNe.  Finally, several have angular
areas of up to 60\,arcmin$^2$ while only the very nearest PNe can attain
such large angular sizes. All the above characteristics are highly  typical
of H{\sc ii} regions.

Yet, despite noting the very attributes that would normally exclude such
objects as being identified as PNe in their paper, these authors gave these 
objects the PN G nomenclature for confirmed PNe.  

It is just as important  
to delete such obvious contaminants from the on-line PN catalogues,
based on improved understanding of their multi-wavelength characteristics,
as it is to augment the literature by new members of the class. Therefore,
we have included an appendix to
the current paper which lists those objects we have indentified in this
study that should be removed
from the extant catalogues and from SIMBAD  as {\it bona fide} PNe.  This
list includes all 14 objects rejected from the Kwok et al.  (2008) sample
(being for the most part H{\sc ii} regions). We also urge the use of PN G
names solely for confirmed PNe, following the accepted
IAU convention. Note that our appendix rejects 25 objects, 16\,percent of the 161
nebulae originally claimed as PNe in our observed area, the same 
proportion of non-PNe as Frew (2008) found for the local volume.

\section{The clean sample of PNe}
After the culls and identity clarifications described above, we
are left with 65 MASH-I, 30 MASH-II, and 41 PNe known prior to MASH,
for a total of 136 PNe in this panchromatic study. 
Note that the 5\,percent fraction of MASH PNe culled using our newly 
developed criteria, that now incorporate GLIMPSE MIR data, is lower than 
the 45\,percent culled from the previously known PNe in our survey region. 
This reflects the care with which MASH catalogues were compiled and 
the heterogeneous nature of the pre-MASH catalogues which had not been 
subjected to our new discriminatory techniques as described by  Frew 
\& (Parker 2010). Note also that none of these culled MASH objects were
designated as true PNe, rather having  the designation of L (likely) 
or P (possible). We are now in a position to utilize our full armoury of
multi-wavelength diagnostic tools in developing robust PN
identification criteria based solely on MIR colours.  
Our cleaned PN sample has resulted 
from the application of our various diagnostic criteria and  we believe it
has minimal taint from the various kinds of PN mimic.  We now proceed to 
examine their multi-wavelength properties  with the emphasis on the 
predictive power inherent in the GLIMPSE MIR data and how to shape
an identification scheme that can be applied in the absence of the
traditional optical information.

Table~\ref{m12attrib}  summarizes the attributes of MASH PNe. 
 This single table  contains the following information:
Col. (1) -- source name; Col. (2-3) -- J2000 Equatorial coordinates with 
units shown;
Col. (4-5)  -- Galactic coordinates in degrees; Col. (6) -- status of 
the PN as true (T),
likely (L), or possible (P); Col. (7) --  optical dimensions in arcsec; 
and Col. (8) --
optical morphology code, described below. Table~\ref{knattrib} is 
similar but but for PNe known prior to MASH and carries 
both a common name and the PN G designation for each nebula.

Note that the PN designations as T (true), L (likely), or P (possible) 
follow the conventions adopted in MASH
catalogues and encapsulate our current best understanding and
interpretation of the object's nature based on all the currently 
available data. 
Of course such designations are subject to change as improved, new, or
different observational data come to light and it might become necessary
to revise the status for a particular  PN, especially for those
objects with the P designation. PN dimensions for newly 
discovered objects are those determined by the MASH project from their 
SHS imagery and correspond to the main body of a given PN.

\section{Nebular morphology}
A detailed description of the morphological classifications of MASH PNe is
given by Parker et al.  (2006) and was utilized in Paper-I.
We adopt the same basic scheme in the current paper. Note though, that
in the absence of high-resolution CCD images, it is often difficult to
provide definitive morphologies for small angular size PNe.

In this paper we investigate  any correlation between optical
morphologies and their MIR characteristics.
The correlation between highly bipolar PNe and strongly enhanced He and
N abundances is already well-known (e.g. Corradi \& Schwarz 1995) 
but the original link between chemistry and morphology was  
noted by Greig (1967, 1971).   Subsequently Peimbert (1978) and 
Peimbert \& Serrano (1980) defined Type I PNe in terms of threshold 
values of He or N abundance. Throughout this paper, we define Type I 
PNe as having N/O $>$ 0.8 following Kingsburgh \& Barlow (1994).  In
other words, these PNe have experienced conversion, through 
envelope burning, of dredged-up primary carbon into nitrogen following 
the third dredge-up (Kingsburgh \& Barlow 1994). It was found early 
on that most Type I PNe are bipolar (Peimbert1978; Peimbert \& 
Torres-Peimbert 1983), while later surveys of bipolar PNe (Corradi 
\& Schwarz 1995)confirmed their chemical peculiarities and added 
enhanced Ne abundance to these. 

Note however, that not all bipolar PNe have Type~I chemistry.
We have found a strong correlation between morphology and Type~I 
chemistry, with the vast majority of Type~I PNe having a bipolar 
morphology (Frew 2008; cf.Phillips 2005).  However the inverse statement 
is not necessarily true, as many close-binary PNe show some 
evidence for bipolarity, but post-common envelope objects as
a group tend to avoid Type~I chemistries (De Marco 2009).

As a class, Type~I PNe have larger than average diameters and expansion
velocities (Corradi \& Schwarz 1995), hotter and more massive central stars
(Tylenda 1989), smaller Galactic scale heights (e.g. Stanghellini 2000), 
and tend to deviate more from the circular rotation of the Galaxy.  
Their association with more massive progenitors than typical
PNe is widely acknowledged on theoretical grounds too (e.g. Becker \&
Iben 1980; Kingsburgh\& Barlow 1994).  This is consistent with the 
high proportion of new Type~I PNe found by MASH at  lower latitudes, 
and the larger fraction of round PNe found at high latitudes by MASH
and later re-affirmed by the Deep Sky Hunter  team in the Digital 
Sky Survey (Jacoby et al.  2010).

An intriguing correlation between bipolarity and the presence of
2.122-$\mu$m H$_2$ emission in PNe was first noted noted by 
Zuckerman \& Gatley (1988) and verified from a large sample of 
nebulae by Kastner et al.  (1996). ``Gatley's rule''  states that H$_2$
detection confirms the bipolar nature of a PN and represents the remnant
of a pre-PN dusty molecular disk.  Therefore, another objective 
of this paper is to seek quantitative differences between bipolar 
and non-bipolar PNe within the GLIMPSE survey which is already
biased toward low latitude PNe with higher than average mass progenitor
stars.  For example, one might expect that PNe bright in 
the {\it v}=1$-$0 S(1) 2.122-$\mu$m line might also be
bright in the IRAC 4.5-$\mu$m band due to emission from the strong  {\it
v}=0$-$0 S(9) 4.694-$\mu$m line and this should be amenable to test 
by comparison of the observed IRAC colours of bipolar and non-bipolar PNe.

A homogeneous treatment of PN morphologies for the pre-MASH ``known"
objects, that is consistent with the above approach, has not been 
available because many objects from the Strasbourg ESO Catalogue:
hereafter ``SEC'' (Acker et al.  1992, 1996), 

\onecolumn
\begin{table*}
\caption{Attributes of  MASH-I \& MASH-II PNe   \label{m12attrib}}  
\begin{tabular}{llllrlll}
Name&           RAJ2000&     DecJ2000&      GLON&      GLAT& Status& Size&  Morph.\\
              &         &            &      deg&       deg&        &  arcsec& type \\
\hline
PHR1806$-$1956& 18 06 55.3 &  $-$19 56 18 & 10.2111&   0.3433&	T&  61$\times$50& Bams\\
PHR1807$-$1827& 18 07 11.7 &  $-$18 27 54 & 11.5293&   1.0039&	P&  7$\times$6&	 E \\
PHR1813$-$1543& 18 13 29.0 &  $-$15 43 19 & 14.6575&   1.0115&	T&  27$\times$21& Eas\\
PHR1815$-$1457& 18 15 06.5 &  $-$14 57 21 & 15.5185&   1.0342&	P&  9$\times$8& Es\\
PHR1818$-$1526& 18 18 59.2 &  $-$15 26 22 & 15.5378&  $-$0.0195& L& 55$\times$11& B\\
PHR1824$-$1505& 18 24 02.1 &  $-$15 05 33 & 16.4158&  $-$0.9312& T& 30$\times$18& Bps\\
PHR1821$-$1353& 18 21 43.9 &  $-$13 53 13 & 17.2190&   0.1272&	P&  20$\times$6& As\\
PHR1826$-$0953& 18 26 26.1 &  $-$09 53 26 & 21.2911&   0.9803&	T&  54$\times$42&  Bs\\
PHR1831$-$0715& 18 31 17.1 & $-$07 15 23 &  24.1803&   1.1436&    P& 19$\times15$& Ias\\
PHR1831$-$0805& 18 31 19.6 &  $-$08 05 43 & 23.4401&   0.7449&	L&  13$\times$9& Eas\\
PHR1834$-$0824& 18 34 41.6 &  $-$08 24 20 & 23.5513&  $-$0.1362& T& 31$\times$26 & Ea\\
PHR1842$-$0539& 18 42 17.9 &  $-$05 39 13 & 26.8632&  $-$0.5529& L& 90$\times$65& Ias\\
PHR1842$-$0637& 18 42 40.4 & $-$06 37 02 &  26.0437&  $-$1.0733&  L& 14$\times$14& Ias\\ 
PHR1843$-$0541& 18 43 10.4 &  $-$05 41 51 & 26.9222&  $-$0.7630& T& 48$\times$39& B?/E\\
PHR1844$-$0517& 18 44 54.0 &  $-$05 17 36 & 27.4764&  $-$0.9616& L& 122$\times$68& Es!\\
PHR1838$-$0417& 18 38 02.2 &  $-$04 17 24 & 27.5860&   1.0186&	P&  15$\times$13& Em \\
PHR1844$-$0503& 18 44 45.7 &  $-$05 03 54 & 27.6643&  $-$0.8265& T& 35$\times$12& Bm?/EM\\ 
PHR1844$-$0452& 18 44 17.3 &  $-$04 52 56 & 27.7721&  $-$0.6350& L& 38$\times$38& Rar\\
PHR1845$-$0343& 18 45 06.0 &  $-$03 43 33 & 28.8931&  $-$0.2907& L& 51$\times$30& B \\
PHR1843$-$0325& 18 43 15.3 &  $-$03 25 27 & 28.9519&   0.2570&	 P& 10$\times$9& Ea\\
PHR1842$-$0246& 18 42 57.1 &  $-$02 46 01 & 29.5024&   0.6246&	 L& 24$\times$13& Em\\
PHR1843$-$0232& 18 43 56.9 &  $-$02 32 08 & 29.8197&   0.5073&	 T& 61$\times$54& Ear\\ 
PHR1846$-$0233& 18 46 02.7 &  $-$02 33 09 & 30.0485&   0.0357&	 T& 36$\times$31& Ear\\
PHR1847$-$0215& 18 47 47.4 &  $-$02 15 30 & 30.5060&  $-$0.2200& T& 20$\times$14& Bs\\
PHR1850$-$0021& 18 50 51.1 & $-$00 21 30 & 32.5462&   $-$0.0330& T& 15$\times$11& B\\ 
PHR1856+0028&   18 56 51.1 &    +00 28 53 & 33.9770&  $-$0.9860&    L& 8$\times$7&  Er\\
PHR1857+0207&   18 57 59.5 &    +02 07 07 & 35.5650&  $-$0.4910&    L& 11$\times$11&  Ea\\
PHR1150$-$6226& 11 50 07.0 &  $-$62 26 32 & 295.9050&  $-$0.4110&   L& 89$\times$77&  R\\
PHR1152$-$6234& 11 52 55.7 &  $-$62 34 10 & 296.2510&  $-$0.4580&   L& 27$\times$24&  R\\
PHR1157$-$6312& 11 57 03.2 &  $-$63 12 44 & 296.8490&  $-$0.9840&   L& 15$\times$13&  B?\\
PHR1202$-$6112& 12 02 18.0 & $-$61 12 47 &  297.0049& 1.0958&     T& 14$\times$13& Es\\ 
PHR1206$-$6122& 12 06 25.5 &  $-$61 22 44 & 297.5680&   1.0230&	  T& 2$\times$11&  E\\
PHR1218$-$6245& 12 18 00.9 &  $-$62 45 38 & 299.1190&  $-$0.1360&   L& 42$\times$31&  Es\\
PHR1219$-$6347& 12 19 08.2 & $-$63 47 01 & 299.3720& $-$1.1353&   L& 22$\times$17& Ea\\
PHR1220$-$6134& 12 20 08.8 & $-$61 34 16 & 299.2160&  1.0747&     L& 10$\times$9& Es\\ 
PHR1223$-$6236& 12 23 58.0 &  $-$62 36 21 & 299.7780&   0.0980&	  T& 48$\times$43&  E\\
PHR1244$-$6231& 12 44 28.5 &  $-$62 31 19 & 302.1330&   0.3510&	  T& 300$\times$235& B \\ 
PHR1246$-$6324& 12 46 26.5 &  $-$63 24 28 & 302.3730&  $-$0.5390&   T& 31$\times$19&  B\\
PHR1250$-$6346& 12 50 04.4 &  $-$63 46 52 & 302.7840&  $-$0.9080&   T& 83$\times$74&  Ea\\
PHR1255$-$6251& 12 55 18.0 &  $-$62 51 04 & 303.3725&   0.0173&	  L& 185$\times$81&  B\\
PHR1257$-$6216& 12 57 51.3 &  $-$62 16 12 & 303.6783&   0.5923&	  P& 19$\times$13&  E\\
PHR1408$-$6229& 14 08 47.3 &  $-$62 29 58 & 311.7300&  $-$0.9500&   T& 82$\times$46&  B\\
PHR1408$-$6106& 14 08 51.7 &  $-$61 06 27 & 312.1525&   0.3741&	  T& 307$\times$264& Es\\
PHR1429$-$6043& 14 29 52.8 &  $-$60 43 57 & 314.6780&  $-$0.1290&   P& 167$\times$131& E\\
PHR1429$-$6003& 14 29 43.9 &  $-$60 03 17 & 314.9220&   0.5180&	  L& 141$\times$110&  E\\
PHR1432$-$6138& 14 32 05.0 & $-$61 38 42 & 314.5840& $-$1.0714&   T& 180$\times$145& Es\\
PHR1437$-$5949& 14 37 53.2 &  $-$59 49 25 & 315.9480&   0.3320&	  T& 103$\times$63&  Ba\\
PHR1447$-$5838& 14 47 41.8 &  $-$58 38 41 & 317.5785&   0.8845&	  P& 57$\times$65&  I\\ 
PHR1457$-$5812& 14 57 35.8 &  $-$58 12 09 & 318.9300&   0.6930&	  T& 31$\times$25&  A\\
PHR1507$-$5925& 15 07 50.2 &  $-$59 25 14 & 319.5050&  $-$1.0140&   T& 22$\times$17&  Ea\\
PHR1544$-$5607& 15 44 56.7 &  $-$56 07 07 & 325.4480&  $-$1.0270&   P& 14$\times$10&  E\\
PHR1552$-$5254& 15 52 56.8 &  $-$52 54 12 & 328.3570&   0.7670&	  T& 31$\times$26&  Es\\
PHR1603$-$5402& 16 03 41.4&   $-$54 02 04 & 328.8410& $-$1.1311&   L& 40$\times21$& A\\
PHR1610$-$5130& 16 10 21.1 &  $-$51 30 54 & 331.2780&   0.0600&	  P& 20$\times$11&  Es\\
PHR1619$-$5131& 16 19 57.6 &  $-$51 31 48 & 332.3493&  $-$0.9814&   P& 11$\times$11&  E\\
PHR1622$-$5038& 16 22 40.6 &  $-$50 38 42 & 333.2746&  $-$0.6547&   L& 21$\times$19&  Ear\\
PHR1619$-$4914& 16 19 40.1 &  $-$49 14 00 & 333.9279&   0.6858&	  T& 36$\times$32&  Rs\\ 
PHR1619$-$4906& 16 19 50.1 &  $-$49 06 52 & 334.0350&   0.7560&	  T& 48$\times$47&  Ra\\
PHR1633$-$4650&  16 33 58.0 &  $-$46 50 07 & 337.3141&   0.6361&	  T& 24$\times$8&  B\\
PHR1635$-$4654& 16 35 51.9 &  $-$46 54 10 & 337.4831&   0.3524&	  P& 77$\times$34& Ia/B?\\
PHR1634$-$4628& 16 34 51.2 &  $-$46 28 28 & 337.6825&   0.7684&	  T& 22$\times$17&  E\\
PHR1639$-$4516& 16 39 22.3 &  $-$45 16 35 & 339.0980&   0.9880&	  T& 38$\times$25&  Er\\
\end{tabular}
\end{table*}
\clearpage

\begin{table*}
{\bf Table~\ref{m12attrib} continued}\\
\begin{tabular}{llllrlll}
Name&           RAJ2000&     DecJ2000&      GLON&      GLAT& Status& Size&  Morph.\\
              &         &            &      deg&       deg&        &  arcsec& type \\
\hline
PHR1646$-$4402& 16 46 27.6 &  $-$44 02 25 & 340.8600&   0.8500&	  L& 71$\times$72&  A\\
PHR1709$-$3931& 17 09 10.8 &  $-$39 31 06 & 347.0320&   0.3500&	  T& 51$\times$13&  B\\
PHR1714$-$4006& 17 14 49.3 &  $-$40 06 09 & 347.2000&  $-$0.8720& T& 20$\times$11&  B?\\
MPA1157$-$6226& 11 57 47.52&  $-$62 26 22 & 296.7712&  $-$0.2110& T& 12$\times$5&   E\\
MPA1235$-$6318& 12 35 21.6 &  $-$63 18 01 & 301.1267&  $-$0.4848& T&  7$\times$5& Eas\\
MPA1315$-$6338& 13 15 30.2 &  $-$63 38 43 & 305.5994&  $-$0.8984& T&  6$\times$6& R\\
BMP1322$-$6330& 13 22 55.4 &  $-$63 30 34 & 306.4344&  $-$0.8520& T& 13$\times$9&  B\\
MPA1324$-$6320& 13 24 16.8 &  $-$63 20 06 & 306.6067&  $-$0.6978& T& 10$\times$10& Rr\\
BMP1329$-$6150& 13 29 50.6 &  $-$61 50 39 & 307.4464&	  0.6908& T& 87$\times$72& Eas\\ 
MPA1337$-$6258& 13 37 55.0 &  $-$62 58 54 & 308.1817&  $-$0.5847& T&  8$\times$7&  E\\
MPA1441$-$6114& 14 41 32.2 &  $-$61 14 18 & 315.7823&  $-$1.1440& T&  7$\times$6& E\\ 
BMP1522$-$5729& 15 22 59.0 &  $-$57 29 59 & 322.1984&  $-$0.4106& T& 13$\times$11& E\\  
MPA1523$-$5710& 15 23 22.5 &  $-$57 10 48 & 322.4178&  $-$0.1720& T& 35$\times$6& Bms\\
BMP1524$-$5746& 15 24 24.0 &  $-$57 46 22 & 322.2075&  $-$0.7427& T& 7 $\times$7& R\\  
MPA1525$-$5528& 15 25 06.1 & $-$55 28 22 & 323.5573&   1.1216&    T& 19$\times$10& B/Ia\\
MPA1605$-$5319& 16 05 37.4 &  $-$53 19 54 & 329.5234&  $-$0.7963& T& 8 $\times$6& E\\   
MPA1618$-$5147& 16 18 42.7 &  $-$51 47 45 & 332.0250&  $-$1.0355& T& 4 $\times$4& S\\
BMP1636$-$4529& 16 36 58.7 &  $-$45 29 29 & 338.6580&     1.1580& T& 11 $\times$9& Em\\
MPA1713$-$4015& 17 13 10.8 &  $-$40 15 56 & 346.8834&  $-$0.7104& T& 7 $\times$5&  E\\
MPA1715$-$3903& 17 15 16.1 &  $-$39 03 49 & 348.0937&  $-$0.3334& T& 78$\times$72& Ra\\
MPA1717$-$3945& 17 17 49.3 & $-$39 45 58 & 347.8068&  $-$1.1426&  T& 6$\times$6& R\\ 
MPA1815$-$1602& 18 15 21.2 &  $-$16 02 56 &  14.5852&  $-$0.4614& L& 6 $\times$5&  Ea\\
MPA1819$-$1307& 18 19 02.3 &  $-$13 07 04 &  17.5886&  $-$1.0682& T& 6 $\times$5&  Ea/Bas?\\
MPA1822$-$1153& 18 22 53.8 &  $-$11 53 10 &  19.1185&  $-$0.8177& T& 11$\times$11&  R\\
MPA1824$-$1126& 18 24 04.1 &  $-$11 26 15 &  19.6493&  $-$0.7741& T& 13$\times$12&  Ear\\
MPA1827$-$1328& 18 27 29.8 &  $-$13 28 19 &  18.2405&  $-$0.9155& L& 12$\times$12&  Rs\\
MPA1832$-$0706& 18 32 22.8 &  $-$07 06 57 &  24.4297&     0.9663& T& 15$\times$13& Emrs\\ 
MPA1840$-$0415& 18 40 25.4 &  $-$04 15 34 &  27.8861&     0.5029& T& 15$\times$6& Em\\ 
MPA1840$-$0529& 18 40 49.2 &  $-$05 29 44 &  26.8326&  $-$0.1517& T& 7 $\times$5&  E\\
MPA1843$-$0556& 18 43 57.8 &  $-$05 56 20 &  26.7958&  $-$1.0497& T& 13$\times$10& E\\
MPA1844$-$0454& 18 44 49.7 &  $-$04 54 00 &  27.8184&  $-$0.7663& T&  7$\times$7& S\\
MPA1851$-$0028& 18 51 47.5 &  $-$00 28 29 &  32.5497&  $-$0.2952& T& 15$\times$12& E\\
MPA1852$-$0044& 18 52 24.2 &  $-$00 44 46 &  32.3778&  $-$0.5549& T& 5 $\times$5&  R\\ 
MPA1852$-$0033& 18 52 25.7 &  $-$00 33 27 &  32.5485&  $-$0.4741& T& 9 $\times$9&  S\\
\hline
\end{tabular} 			    
\end{table*}			    
\clearpage   
\begin{table*}			    
\caption{Attributes of PNe known prior to MASH \label{knattrib}}   
\begin{tabular}{lllllrlll}	    
Name&   PN G&        RAJ2000&     DecJ2000&      GLON&	GLAT& Status& Size&  Morph.\\
              &         &            &      deg&	deg&	    &  arcsec& type \\
\hline				    
Hen\,2-83& 300.2$+$00.6& 12 28 43.9 & $-$62 05 35 & 300.279& 0.662&  T&  4.5$\times$4.5& E\\ 
Hen\,2-84&300.4$-$00.9&  12 28 46.8 & $-$63 44 36 & 300.429& -0.982&  T&  36$\times$24& B\\  
Hen\,2-85&300.5$-$01.1&  12 30 07.7 & $-$63 53 01 & 300.589& $-$1.109& T& 9$\times$8& E \\
Wray16-121& 302.6$-$00.9& 12 48 31.2 & $-$63 50 04 & 302.610& -0.965& T& 65$\times$54& Es   \\ 
TH 2-A& 306.4$-$00.6&  13 22 33.8 & $-$63 21 02 & 306.414& -0.689& T& 27$\times$25& Epr \\ 
WKG2&308.4$+$00.4& 13 38 42.7 & $-$61 55 47 & 308.462& 0.433&  T&  38$\times$32&  Er \\  
 Hen\,2-96&309.0+00.8& 13 42 36.2 & $-$61 22 29 & 309.021& 0.891&  T& 17$\times$14& E  \\ 
... & 313.3+00.3& 14 18 27.8 & -60 47 10 & 313.355& 0.312&  P&  28$\times$28&  R \\  
 Hen\,2-111& 315.0-00.3&  14 33 18.0 & -60 49 37 & 315.030& -0.371&  T&  29$\times$15& Bmp   \\  
GLMP387& 316.2+00.8&  14 38 19.9 & -59 11 46 & 316.247& 0.884&  P&  6.5$\times$5.5& S  \\ 
Pe2-8&322.4-00.1& 15 23 43.0 & -57 09 25 & 322.469& -0.178&  T&  1.9$\times$1.3& S \\ 
GLMP437& 327.8-00.8&  15 57 21.1 & -54 30 46 & 327.829& -0.888&  P&  6.5&  S \\   
Hen\,2-145& 331.4+00.5& 16 08 58.8 & -51 01 58 & 331.448& 0.562&  T&  17$\times$15& B    \\ 
Mz~3&331.7-01.0&  16 17 13.4 & -51 59 11 & 331.727& -1.011&  P&  48$\times$23& B \\  
Hen\,2-169& 335.4-01.1&   16 34 13.2 &  -49 21 13 & 335.492& $-$1.103& T& 22$\times$19&  Bs  \\
H\,1-3& 342.7+00.7&  16 53 31.2 & -42 39 22 & 342.744& 0.752&  T&  20$\times$16& B \\ 
GLMP495& 342.2-00.3&  16 56 34.1 & -43 46 14 & 342.226& -0.380&  P&  7.5& S  \\   
H\,1-5& 343.9+00.8& 16 57 23.8 & -41 37 58 & 343.992& 0.835&  T&  69$\times$5&  E  \\ 
HaTr\,5& 343.3-00.6& 17 01 28.6& -43 05 57& 343.305& -0.664&  T&  120$\times$110& Rar    \\  
IC4637& 345.4+00.1&  17 05 10.6 & -40 53 08 & 345.479& 0.140&  T&  19$\times$14& Eap    \\  
NGC6337& 349.3-01.1&  17 22 15.7 & -38 29 03 & 349.3510& -01.116& T& 48$\times$42& Erspm \\  
NGC6302& 349.5+01.0& 17 13 44.4 & -37 06 16 & 349.5075& $-$1.016& T& 90$\times$35& Bs  \\
NGC6537&010.1+00.7& 18 05 13.0 & -19 50 35 & 10.0987& 0.740&  T&  11$\times$10&  Bmps \\ 
Sab39& 011.7+00.2& 18 10 19.7& -18 39 10& 11.7255& 0.264&  T&  17$\times$12& E \\  
NGC6567&011.7-00.6& 18 13 45.1 & -19 04 35 & 11.7433& -0.650&  T&  8$\times$6&  Ep  \\  
M1-43&011.7+00.0& 18 11 48.9 & -18 46 22 & 11.7898& -0.102&  T&  15$\times$12& S    \\ 
M3-53&019.9+00.9& 18 24 07.9 & -11 06 42 & 19.9447& 0.912&  L&  10$\times$9&  S   \\ 
PM\,1-231& 020.4+00.6& 18 25 58.1 & -10 45 29 & 20.468&  0.679& L&  9$\times$7&  S \\  
PM\,1-235& 021.6+00.8& 18 27 45.6 & -09 38 13 & 21.6656& 0.811&  P&  2.5& S \\  
GLMP781&021.1+00.4& 18 28 01.2 & -10 14 08 & 21.1654& 0.476&  P& 6.5$\times$5.5& S \\  
Abell45& 020.2-00.6& 18 30 15.6 &  -11 37 02 & 20.1965& -0.652&  L&  300$\times$280& Ers   \\ 
Mac\,1-13& 022.5+01.0& 18 28 35.3 &  -08 43 22 & 22.5699& 1.055&  T& 23$\times$17&  E/B  \\
M3-28&021.8-00.4& 18 32 41.3 & -10 05 52 & 21.8197& -0.478&  T&  24$\times$12& Bp  \\ 
M3-55&021.7-00.6& 18 33 14.9 & -10 15 20 & 21.7431&-0.673&  T&   12$\times$9& B \\ 
M1-51&020.9-01.1& 18 33 29.04 &  $-$11 07 27 & 20.9991& -1.125&  T& 15$\times$8& B \\ 
M2-45&027.7+00.7& 18 39 21.8 & -04 19 51 & 27.7017& 0.705& T& 15$\times$12& E    \\ 
Pe\,1-14& 025.9-00.9& 18 42 07.9 & -06 40 55 & 25.9269& -0.984&  T& 13$\times$10&  B \\ 
Abell48& 029.0+00.4& 18 42 46.8 & -03 13 16 & 29.0786& 0.455&  T&  50$\times$43& Em \\  
TDC\,1& 029.2+00.0& 18 44 53.5 & -03 20 33 & 29.2111&-0.069&  L& 6$\times$5& E  \\   
HaTr\,10& 031.3-00.5&  18 50 24.5 & -01 40 19 & 31.3267&-0.534&  T&  34$\times$28& B   \\  
WeSb4& 031.9-00.3& 18 50 40.3 & -01 03 13 & 31.9067&-0.310&  T&   69$\times$42&  B \\ 
CBSS3& 032.9-00.7& 18 54 06.7 & -00 20 02 & 32.9396&-0.748&  T& 7$\times$5& E \\  
GLMP844& 033.4-00.6& 18 54 34.80 &  00 11 04.6 & 33.4544& 0.615 & T&   3.5&  S  \\ 
\hline
\end{tabular} 
\end{table*}
\noindent
 One GLIMPSE detected nebula in Table\,2, G313.3+00.3 (Cohen et al. 2005),  has no common name in col.(2)
because its optical counterpart is too faint to characterize the nebula.
\clearpage

\begin{landscape}
\begin{table*}	
\begin{center}
{\footnotesize 
\caption{Derived data for all our samples of PNe \label{derived}}
\begin{tabular}{lrrrrrrlrrrrrrlll}

Name& {\bf [3.6]}& 1$-$2& 1$-$3& 1$-$4& 2$-$3& 2$-$4& 3$-$4& False&  H$\alpha$/[S{\sc ii}]& H$\alpha$/[N{\sc ii}]& N$_e$([S{\sc ii}]& MOST/& Radio& 8.0\,$\mu$m& MIR/Radio\\   
    &      &      &      &      &      &  &    & colour&    log$_{10}$&   log$_{10}$&  cm$^{-3}$& NVSS&   mJy&   mJy  \\
\hline
PHR1150$-$6226& 9.70& 0.60& 1.81& 2.35& 1.21& 1.74& 0.54&  X& ...& ...& ...& ...& ...& ...& ....\\ 
PHR1152$-$6234& ...&  ...& ...& ...& 1.09& 1.67& 0.58& X& 0.13 & $$-$$0.11& ...& ...&  ...& ...& ... \\
PHR1157$-$6312& 12.88& 1.72& 2.37& 4.42& 0.65& 2.69& 2.05& R& 0.44 &$-$0.35&  700&   m&  6.4& 22& 3.5\\
PHR1202$-$6112& ...& ...& ...&  ...&  ...&  2.84& ...& u& 0.05& $-$0.79&  180&   m& 8.3& 10.6& 1.3\\ 
PHR1206$-$6122& 14.19& 1.50& 2.25& 2.72& 0.75& 1.21& 0.47& V& 0.55& 0.14&  150&  ...&  ...& ...& ... \\
PHR1218$-$6245& ...&  &...& ...& ...& ...& ...& u& 0.28 &  $-$0.28& 130&  ...&  ...& ...\\ %
PHR1219$-$6347& 12.32&  0.14&   2.25&	2.76&   2.11&   2.63&   0.52& O&   ...&   $-$0.16&   ...&     m&  230& 7.8& 0.03\\ 
PHR1220$-$6134& ...&   ...& ...&  ...&  ...& 1.43&  ...& u& 0.88:&    0.05& ...&   m&  9.4&  8.7& 0.9\\ 
PHR1223$-$6236& 11.50&  $-$0.31& 0.34& 1.60& 0.64& 1.89& 1.26& X& 0.53 &$-$0.24& ...& ...& ...& ...& ... \\
PHR1244$-$6231& ...& ...& ...& ...& ..3.34& 4.25& 0.91&  X& 0.18& $-$0.89&  100& ...& ...& ...& ... \\ 
PHR1246$-$6324&  11.05& 0.63& 2.49& 4.27& 1.85& 3.64& 1.79&  R& 0.90& 0.10&  850&  m&  14& 100& 7.4\\ 
PHR1250$-$6346& ...& ...& ...& ...& 1.46& 3.29& 1.85& X& 0.37 &$-$0.26& 150& ...&  ...& ...& ... \\ 
PHR1255$-$6251&  ...& ...& ...& ...& 0.51 &$-$1.68 &$-$2.18& X& $-$0.13 &$-$1.18& ...& ...& ...& ...& ... \\ 
PHR1257$-$6216& 12.24&   0.21& 0.82& 2.14& 0.61& 1.94& 1.33& V& ...&  ...& ...&  ...& ...& ...& ... \\ 
PHR1408$-$6106& 9.82& ...& ...& ...& ...& ...& ...& u& 0.28 & $-$0.28& 130&  ...&  ...& ...& ...\\ 
PHR1408$-$6229& 13.19&  2.44& ...& ...& ...& ...& ...& u& 0.27 &$-$0.75& ...& ...& ...& ...& ... \\ 
PHR1429$-$6003& 9.53& 1.63& 3.02& 4.39& 1.42& 2.78& 1.37& X& $-$0.16 &$-$0.71&  10& ...& ...& ...& ... \\ 
PHR1429$-$6043& 8.46&  0.68& 2.73& 4.25& 2.04& 3.56& 1.53& X& ... & ...& ...&  ...& ...& ...& ... \\
PHR1432$-$6138& 9.03&   $-$1.12&   1.46&	4.70&   2.58&   5.82&   3.24& X& 0.23&   $-$0.43&  270& ...& ... ...& ...& \\ 
PHR1437$-$5949& 9.21& 0.78& 0.89& 3.23& 0.11& 2.44& 2.34&  R& 0.03 &$-$0.91&  ...&    m&  12& 240& 20\\ 
PHR1447$-$5838& ...&  ...& ...& ...& ...& 3.33& ...& u& 0.41& 0.07&  270& ...&  ...& ...& ... \\ 
PHR1457$-$5812& 10.48&  0.20& 1.29& 3.75& 1.09& 3.54& 2.46&  R& 0.46 &$-$0.55& 1500&       m&  80& 120& 1.5\\ 
PHR1507$-$5925& 10.56&  0.88& 1.74& 3.01& 0.86& 2.12& 1.27&  V& 0.83& 0.45& 900&     m&  24& 74& 3.6\\ 
PHR1544$-$5607& ...&  ...& ...& ...& 0.54& 3.34& 2.81& X& $-$0.07: &$-$0.12:& ...& ...&  ...& ...& ... \\ 
PHR1552$-$5254& 10.60& 0.65& 1.38& 3.80& 0.74& 3.14& 2.42&  R& 0.37 &$-$0.41& 600&    m&  21& 100& 4.4\\ 
PHR1603$-$5402& ...&   ...& ...&  ...&  ...&  3.41&  ...& u&  0.39& 0.39& ...&  ...& ...&  ...& ... \\
PHR1610$-$5130& 13.19&  0.17& ...& ...& ...& ...& ...& u& 0.51& 0.08& 400& ...&  ...& ...& ... \\ %
PHR1619$-$4906& ...&  ...& ...& ...& 2.39& 4.43&  2.05& X& ....& $-$0.63& ...& ...& ...& ...& ... \\ 
PHR1619$-$4914& 8.78&  0.93& 1.02& 3.14& 0.10& 2.22& 2.12& R& 0.87&  $-$0.41&  1300&   m&  240& 720& 3.0\\ 
PHR1619$-$5131& 11.51&   0.18& .2.44& 4.15& 3.16& 4.95& 1.80& V& ... &... & ...&    m&  20& 73& 3.6\\
PHR1622$-$5038& 11.13&  1.69& 2.46& 4.01& 0.77& 2.31& 1.55&  V& 0.52& 0.24& ...& ...&  ...& ...& ... \\ 
PHR1633$-$4650& ...&  ...& ...& ...& 2.74& 4.62& 1.90& O& 0.03 &$-$0.88& 400&   m&  27& 47& 1.7\\ 
PHR1635$-$4654& ...& ...&  ...&  ...&  ...&  ...&  u&  ...& ...& ...& ...& ...& ...  \\ 
PHR1634$-$4628& 11.77&   1.06& 2.09& 3.42& 1.03& 2.35& 1.33& X& 0.17 &$-$0.19& 60& ...& ...& ...& ... \\
PHR1639$-$4516& 9.62&  1.52&  $-$0.07& ...&  1.11&...& ...& u& 0.49 &$-$0.31& 70& ...& ...& ...& ...\\ 
PHR1646$-$4402& 10.21&   $-$0.54& 0.47& 1.51& 1.00& 2.03& 1.04& X&...& $-$0.64& ...& ...& ...& ...& ...  \\
PHR1709$-$3931& 14.70&  2.89& 6.57& 7.24& 3.68& 4.34& 0.67& X& 0.02 &$-$0.88& ...& ...& ...& ...& ... \\ 
PHR1714$-$4006& 12.59&  1.47& 3.09& 3.93& 1.62& 2.45& 0.84&  R& 0.52 &$-$0.46& ...&    m&  9.7& 20& 2.1\\ 
PHR1806$-$1956& 11.55& ...& 3.47& 4.67& ...& ...& 2.73&  u& 0.28 &$-$0.50&  700& ...& ...& ...& ... \\ 
PHR1807$-$1827& 17.13& 3.38& 4.74& 6.92& 1.38& 3.54& 2.17&  X& .... & 0.11& ...& ...& ...& ...& ... \\ 
PHR1813$-$1543& ...& ...& ...& ...& ...& ...& 1.77&  u& 0.65 & 0.19& ...&...& ...& ...& ...  \\ 
PHR1815$-$1457& ...& ...& ...& ...& ...& 3.18& ...& u&....  &.... &   ...&         n& 9.2& 12& 1.3\\ 
PHR1818$-$1526& 9.80& 1.53& 3.37& 1.41& 1.84 &$-$0.13 &$-$1.96& X& 0.20 &$-$0.70& 250&...& ...& ...& ... \\ 
PHR1824$-$1505& 12.94& 1.57& 3.10& 1.19& 1.53& $-$0.39& $-$1.91& X& 0.46: &$-$0.50& 80:& ...& ...& ...& ... \\
PHR1821$-$1353& 11.36& ...&   ...&   ...&  ...&   ...&  ...&   u&    &.... &   ...&    ...&  ...&...& 1.3\\
PHR1826$-$0953& 9.77& 0.56& 2.45& 3.92& 1.66& 3.36& 1.46&  X& 0.33 &$-$0.33& 170&  n&  8.4& 130& 15\\
PHR1831$-$0715& ...&  ...& ...&  ...&  ...&  2.14&  ...&  u&  ...&  0.29&   ...&     n&  5.9& 21& 3.6\\ 
\end{tabular}
}
\end{center}
\end{table*}
\end{landscape}
\clearpage 

\begin{landscape}
\begin{table*}
\begin{center}			   
{\bf Table~\ref{derived} continued}\\
{\footnotesize
\begin{tabular}{lrrrrrrlrrrrrrlll}
Name& {\bf[3.6]}& 1$-$2& 1$-$3& 1$-$4& 2$-$3& 2$-$4& 3$-$4& False&  H$\alpha$/[S{\sc ii}]& H$\alpha$/[N{\sc ii}]& N$_e$([S{\sc ii}]& MOST/& 8.0\,$\mu$m& Radio& MIR/Radio\\   
    &      &      &      &      &  &     &      & colour&    log$_{10}$&   log$_{10}$&  cm$^{-3}$& NVSS&   mJy&   mJy  \\     
\hline
PHR1831$-$0805& 12.20& 0.50& 1.93& 3.70& 1.43& 3.18& 1.76& V& 0.25&  $-$0.49&   2800& ...&  ...& ...&   ...    \\
PHR1834$-$0824& 11.95& ...& ...&  ...&  ...&  ...&  ...&  u&  ...& ...& ...& ...& ...& ...  \\
PHR1838$-$0417& ...& ...&   ...&  ...& $-$0.06& 1.99& 2.67& V&  ...&   1.15&    ...&   ...&  ...& ...& ...   \\ 
PHR1842$-$0246& 10.55& 0.36& $-$2.23& $-$0.15& $-$1.98& ...& 2.08&   R& ....  &$-$0.58&  ...& ...& ...& ... \\ 
PHR1842$-$0539& 8.52& 0.29& 1.19& 2.41& 1.29& 2.51& 1.23& X& $-$0.26 &$-$1.25&  1000& ...&  ...&...& ... \\
PHR1842$-$0637& 10.36& $-$0.08& $-$0.07&  0.61&  0.01&  0.69&  0.68& V&   $-$0.84&   ...& ...&  ...& ...&  ...&  \\ 
PHR1843$-$0232& 8.46&  1.13& 3.18& 5.02& 2.05& 3.88& 1.84& X& 0.50& $-$0.18& ...& n& 33& 700&  21 \\
PHR1843$-$0325& 11.30&  1.10& 2.71& 4.66& 1.60& 3.55& 1.94& V&.... & 0.38&   ...&   n&  16& 160& 10\\ 
PHR1843$-$0541& ...& ...& ...& ...&$-$1.04& 0.00& 1.09& X&  0.09 &$-$0.90&  400& ...& ...&...&  ...  \\ 
PHR1844$-$0452& 10.20& 0.58& 1.71& 1.34& 1.17& 0.78& 0.38& X& 0.29 &$-$0.75& 500& ...& ...& ...&  ... \\
PHR1844$-$0503& 12.06& 0.80& 1.86& 2.98& 1.05& 2.20& 1.38& O& 0.81& $-$0.36&   2500&   n&  3.2& 58& 18\\ 
PHR1844$-$0517& 11.49& $-$2.08& 0.54& 2.42& 2.62& 4.48& 1.88& X& 0.47 &$-$0.66&  ...& ...& ...& ... \\ 
PHR1845$-$0343& 9.16& 0.64& 1.07& 2.70& 1.10& 3.01& 1.91&O& ....  &$-$0.84& ...&  ...& ...& ...&  ...\\ 
PHR1846$-$0233& ...& ...& ...& .... &...&  1.67& ...& u&.... & $-$0.38& ...& ...&  ...&...&  ...  \\ 
PHR1847$-$0215& 10.40& ...& 1.69& 3.95& 5.26& 7.51& 2.21&  V& 0.57& $-$0.61& ...& ...&  ...& ...&... \\ 
PHR1850$-$0021& 12.83&  0.88& 1.03&  2.99&  0.18&  2.15&  1.97& X&   $-$0.42& ...&   ...& n&  5.0& 7.4& 1.5\\
PHR1856+0028& 11.83&  0.21& 0.74& 2.64& 0.35& 2.17& 1.91&  V& 0.61 &$-$0.04& 750& ...&  ...& ...& ... \\ 
PHR1857+0207& 10.40&  0.67& 1.47& 3.48& 0.81& 2.82& 2.00& R& .... & 0.51&  ...&  n&  100& 130& 1.3 \\ 
MPA1157$-$6226& 12.72&  0.71& 1.37& 2.25& 0.66& 1.55& 0.89& R& 0.46 &$-$0.73& 300& ...&  ...& ...& ...\\
MPA1235$-$6318& 13.91& 1.60& 2.06& 3.93& 0.46& 2.34& 1.88& O& 0.32 &$-$0.49&  650& ...& ...& ...& ... \\ 
MPA1315$-$6338& 11.35& 0.38& 1.24& 2.83& 0.86& 2.45& 1.59& O& 0.52 &$-$0.23&   630&   m&  4.9& 25& 5.0\\ 
BMP1322$-$6330& 13.36& 0.31& 1.53& 3.21& 1.22& 2.90& 1.67& O& $-$0.11 &$-$1.15&  90&  m& ...& ...& ... \\ 
MPA1324$-$6320& 13.32& 0.61& 1.06& 2.83& 0.45& 2.22& 1.77&  R& 0.37 &$-$0.24&  500&  m&  ...& ...& ... \\
BMP1329$-$6150& 11.39&  1.67& 2.24& 4.74& 0.57& 3.07& 2.50& X& ....  &$-$0.48& ...&  m& ...& ...& ... \\
MPA1337$-$6258& 12.20& 0.69& 1.11& 2.50& 0.42& 1.81& 1.39&  R& 0.93: &$-$0.31&  800:& ...& ...& ...& ... \\ 
MPA1441$-$6114& 11.32& 0.95& 1.13& 2.85& 0.17& 1.89& 1.72&  R& 2.06:& 1.31:&   60&     m& 20& 26& 1.3 \\ 
BMP1522$-$5729& 11.10& 1.25& 1.78& 3.99& 0.53& 2.74& 2.21&  R& 0.85 &$-$0.35&  3600&   m&  17& 90& 5.2\\
MPA1523$-$5710& 12.18& 1.78& 2.25& 4.14& 0.48& 2.36& 1.88&  R& 0.47 &$-$0.66&  1300&  m&  27& 38& 1.4\\ 
BMP1524$-$5746& 9.63& 0.84& 1.35& 3.17& 0.52& 2.33& 1.82&  R& 0.71 &$-$0.19& 2100&  m&  27& 170& 6.1\\ 
MPA1605$-$5319& 11.78& 0.87& 1.90& 3.75& 1.03& 2.88& 1.85&  O& 0.92 &$-$0.05& 2000&    m&  5.1& 39& 7.6\\ 
MPA1618$-$5147& 12.60& 0.39& 0.26& 1.86& $-$0.13 &  1.47& 1.61& V& 0.84 &$-$0.34&  470& ...& ...& ...& ...\\ 
BMP1636$-$4529& ...& ...& ...& ...& 2.83& ...& ...& u& 0.97& $-$0.18& 1300&  m& 22&  60& 2.7\\  
MPA1713$-$4015& 11.12& 0.69& 2.02& 4.06& 1.34& 3.37& 2.04&  O& 0.83 &$-$0.18& 2000& ...&  ...& ...& ...\\ 
MPA1715$-$3903& 9.67& 2.75& 1.95& 3.70 & $-$0.80&  0.95& 1.75& X& 0.47 &$-$0.17:& 300& ...& ...& ...& ... \\ 
MPA1717$-$3945& 12.41& ...& 1.29&  ...&  ...&  ...&  ...& u& ...&   1.09&   ...&      m&  ...& ...& ...\\  
MPA1815$-$1602& 11.67& 1.61& 2.09& 4.08& 0.48& 2.47& 1.99& O& 0.43 &$-$0.53& 1400&   n&  11& 28& 2.6\\  
MPA1819$-$1307& 13.79& 1.95&  1.70& 4.51&  $-$0.13&  2.56& 2.82& R&.... & 1.20&  ...&   n&  5.2& 12& 2.4\\  
MPA1822$-$1153& 11.86&  0.70& 0.75& 2.68& 0.06& 1.98& 2.68& R& .... & .... &  ...& ...& ...  \\  
MPA1824$-$1126& 11.11& 0.28&  $-$0.62& 1.32 & $-$0.91&  1.04& 1.94& V&  1.10:& 0.88& 700:& ...& ...& ...& ... \\ 
MPA1827$-$1328& 10.77& 1.86& 1.46& 3.09 & $-$0.40&  1.23& 1.63& R&  0.41:& 0.24?& ...& n&  34& 54& 1.6\\  
MPA1832$-$0706& 10.17& 0.28& 2.01& 3.83& 1.72& 3.55& 1.83&  O& 0.88& $$-$$0.10& 1100&    n&  5.5& 210& 38\\  
MPA1840$-$0415& 12.38& 0.81& 1.04& 2.34& 0.23& 1.52& 1.30& R&  0.42 &$-$0.73&  4500& ...& ...& ...& ... \\ 
MPA1840$-$0529& 12.03& 0.90& 1.03& 2.98& 0.13& 2.08& 1.96&  V& 0.26: &$-$0.48& 800:&    n&  1.4& 15& 11\\ 
MPA1843$-$0556& 11.40& 0.32& 0.20& 2.17 & $-$0.12&  1.84& 1.97& O&  0.40 &$-$0.50&   1400&      n&  5.5& 13& 2.3\\  
MPA1844$-$0454& 12.68& 0.74& 0.84& 2.84& 0.10& 2.10& 2.00& O& 0.82: &$-$0.59&  2400:& ...&  ...& ...& ... \\  
MPA1851$-$0028& 12.23&  1.29& 1.39& 3.80& 0.10& 2.51& 2.41& O& 0.35 &$-$0.47& 140& ...& ...& ...& ...\\  
MPA1852$-$0044& 13.59& 0.68& 1.85& 3.86& 1.17& 3.19& 2.01& R& ...& $-$0.47&  ...& ...& ...& ...&... \\  
MPA1852$-$0033& 9.87& 0.69& 1.78& 3.81& 1.09& 3.12& 2.03& O& 1.37&    0.12& 24000&    n&  20& 240& 12\\ 
\end{tabular}           
}   
\end{center}
\end{table*}
\end{landscape}
\clearpage

\begin{landscape}
\begin{table*}
\begin{center}			   
{\bf Table~\ref{derived} continued}\\
{\footnotesize 
\begin{tabular}{lrrrrrrlrrrrrrlll}       
Name & {\bf[3.6]}    &1$-$2& 1$-$3& 1$-$4& 2$-$3& 2$-$4& 3$-$4& False&  H$\alpha$/[S{\sc ii}]& H$\alpha$/[N{\sc ii}]&  N$_e$([S{\sc ii}]& MOST/& Radio& 8.0$\mu$m& MIR/Radio\\
    &      &      &      &      &  &    &      & colour&      log$_{10}$&   log$_{10}$&  cm$^{-3}$& NVSS&   mJy&   mJy  \\     
\hline
Hen\,2-83& 9.84&  0.91& 1.56& 3.87& 0.65& 2.96& 2.31& R& 1.23& $-$0.16&   7000&   m&  41& 260& 6.4\\  
Hen\,2-84& 11.15  &1.31& 2.07& 2.94& 0.76& 1.63& 0.87& R& 0.44&   $-$0.38&   850&      m& 12& 33& 2.8 \\ 
He\, 2-85& 10.18&  1.61& 1.06&  2.90& $-$0.55&  1.29&  1.83& R& 1.32&    0.54& 3400&     m&  77& 78& 1.0 \\ 
Wray16-121& 13.08& 0.24& 1.33& 3.52& 1.09& 3.28& 2.19& X& 0.24&   $-$0.73&  1100&      m&  27& 300& 11\\ 
TH 2-A& 9.26& $-$0.13& 1.12& 3.26& 1.25& 3.39& 2.14& R& 1.21&    0.26& 1000&     m&  16& 250& 16 \\ 
WKG2& 11.45& 1.00& 1.67& 4.19& 0.67& 3.18& 2.51& X&  0.87& $-$0.29&  140& ...& ...& ...& ... \\ 
Hen\,2-96& 9.17&  0.52& 2.11& 4.12& 1.58& 3.59& 2.01& O& 1.58&    0.47&   8300&     m&  17& 600& 35\\ 
Hen\,2-111& 7.68& 2.20& 3.49& 25.82& 1.29& 3.62& 0.63& O& 0.26&   $-$0.78&   700&       m&  82& 870& 11 \\  
GLMP387& 9.18& 0.82& 2.07& 4.21& 1.25& 3.39& 2.14& O& ...& ...& ...&     m&  13& 650& 49\\ 
Pe2-8& 8.27& 1.08& 1.90& 4.34& 0.83& 3.26& 2.43& R&  2.12&    0.66&  11000&     m&  27& 1700&62 \\  
GLMP437& 9.75& 0.61& 1.76& 3.81& 1.15& 3.20& 2.05& O& ...& ...& ...&    m& 62& 260& 4.2\\  
Hen\,2-145& 9.82&  0.52& 3.15& 4.84& 2.62& 4.31& 1.69& X& ...&  $-$0.37& ...&  m&  58& 640& 11\\  
Hen\,2-169& 10.26&   1.21& 2.06&  4.37&  0.85&  3.16&  2.32& O&  0.28&   $-$0.76&   1400&     m& 97& 280& 2.9 \\  
H\,1-3& 9.32& 0.11& 0.99& 2.57& 0.88& 2.45& 1.57& R& 0.28&   $-$0.54&  500&     m&  32& 120& 3.9\\  
GLMP495& 9.30&  1.26& 1.16& 2.91 &-0.10& 1.64& 1.75&  R& 1.31&   0.60& 3900&    m&  76& 180& 2.3\\  
H\,1-5& 8.64& 0.99& 1.53& 3.70& 0.55& 2.72& 2.17& R&1.26&    0.29&   24000&       m&  115& 670& 5.8\\  
HaTr\,5& 8.69&  $-$0.14& 1.89& 4.21& 2.03& 4.35& 2.32& X& ...& ...& ...&  ...& ...& ...\\ 
IC4637& 8.67&  0.84& 1.52& 2.47& 0.68& 1.63& 0.95& O& ...&    1.23&   ...&    m&  210& 210& 1.0\\
NGC6337& ...& ...& & ...& ...& ...& ...& u&  0.22& $-$0.80& 260&  m&  106& ...& ... \\ 
NGC6302& 6.04&   1.59& 2.56&  4.82&  0.97&  3.23&  2.26& O&  0.90&   $-$0.36&   5400&    m&1200& 20000& 16\\ 
NGC6537&  7.33&  1.29& 2.35& 4.40& 1.06& 3.11& 2.05& O& 0.87&   $-$0.14&   ...&   n&  430& 4300& 10\\ 
Sab39& 12.08& 2.40& 0.88& 3.86 &-1.51& 1.46& 2.98& R&  0.82&    0.04&   1100&  n&  13& 33& 2.5\\ 
NGC6567& 8.03& 0.70&  1.76& 4.10& 1.07& 3.41& 2.34& O&  2.28&    1.36&   11000&  n&  160& 1700& 10\\ 
M1-43& 9.94& 0.63& $-$0.21& 1.95& $-$0.84& 1.32& 2.16&  V& ...&    1.00& ...& ...& ...& ...& ...... \\
M3-53& 10.55& 0.75& 1.76& 3.87& 1.00& 3.11& 2.11& O& 0.68&   $-$0.32&   3200&   n&  20& 130& 6.9\\
PM\,1-231& 10.40& 1.23& 1.69& 4.08& 0.46& 2.85& 2.39& R& 1.25:&    0.09&   3900:&   n&  49& 190& 3.9\\ 
PM\,1-235& 9.90& 1.23& 2.67& 4.54& 1.44& 3.31& 1.87& O& ...& ...& ...& n&  16& 450& 28 \\ 
GLMP781& 8.18& 0.84& 2.17& 4.28& 1.34& 3.45& 2.11& R& ...& ...& ...&n& 38& 1700& 46 \\ 
Abell45& ...& ...& ...& ...& 2.30 &...& ...& u& ...& ...& ...& ...& ...& ...& ... \\  
Mac\,1-13& 10.82&  1.29& 1.46&  3.61&  0.17&  2.32&  2.15& R& 0.33&   $-$0.40&  1100&      n&  30& 82& 2.7 \\ 
M3-28& 9.13& 0.81& 2.36& 4.05& 1.56& 3.24& 1.68& R& 1.12&   $-$0.34&  1400&    n&  20& 590& 30\\  
M3-55& 11.32&  $-$0.03& 1.52& 3.27& 1.55& 3.30& 1.75& R&  0.60&  $-$0.50& 1700& ...& ...& ...& ... \\ 
M1-51& 8.06&  ...& 0.58&  ...&  ...&  ...&  ...& u&  1.12&   $-$0.24&  11000&   n& 250&...& ...\\  
M2-45& 9.64& 1.02& 0.18& 3.27 &-0.84& 2.24& 3.09& V& 1.24&    0.39&   3200&    n&  110& 180& 1.7\\  
Pe\,1-14& 12.52&  1.34& 2.02& 3.68& 0.68& 2.34& 1.66& R& 0.46&   $-$0.42&   1150&    n&  2.2& 18& 8.4\\ 
Abell48& 9.55& 1.59& 0.67& 3.50 &-0.92& 1.92& 2.83& R& 1.32&    0.58&   240&       n&  160& 240& 1.5\\ 
TDC\,1& 9.41& 0.93& 1.73& 3.96& 0.80& 3.03& 2.23& R& ...& ...& ...&  n&  140& 420& 3.0\\ 
HaTr\,10& 11.11& 0.51& 0.93& 1.95& 0.42& 1.45& 1.03& V&  0.22&   $-$0.80&  260& ...& ...& ...& ... \\ 
WeSb4& 10.04& 1.15& 1.30& 2.82& 0.15& 1.67& 1.53& V&  0.20&   $-$0.90&   550&     n&  4.3& 82& 19\\  
CBSS3& 11.86& 0.69& 1.38& 3.13& 0.69& 2.44& 1.75& O& 0.35&   $-$0.49&   900& n&  2.0& 20& 810 \\ 
GLMP844& 8.71& 0.79& 2.35& 4.40& 1.56& 3.61& 2.05& R& ...&   $-$0.10:&   ...&   n& 6.5& 1200& 180 \\ 

\hline  
\hline& 

\end{tabular}  
}
\end{center}
\end{table*}
\end{landscape}
\clearpage

\twocolumn
\noindent
 \twocolumn

\onecolumn
\begin{figure}
\vspace{22cm}  
\includegraphics{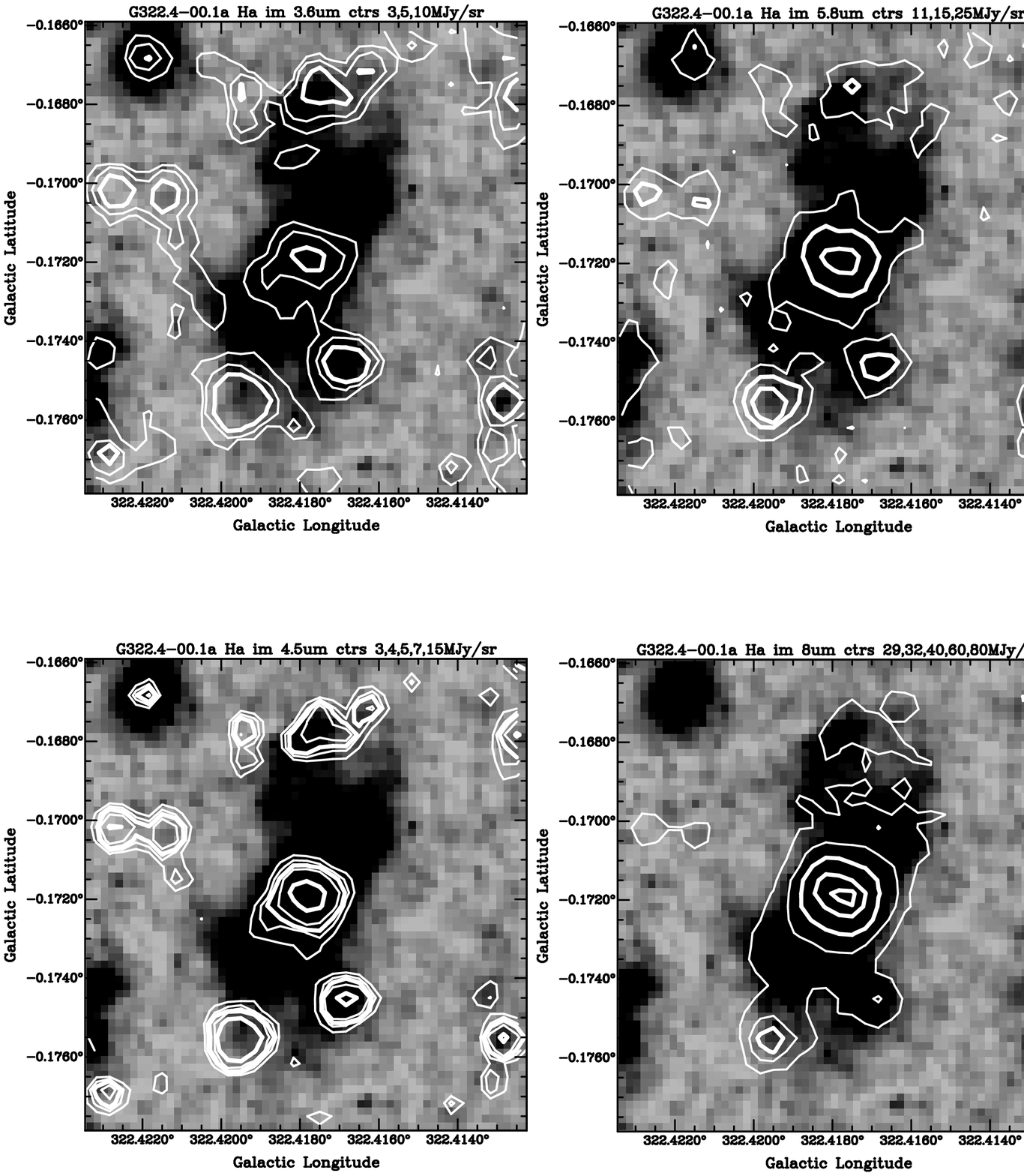}   
\caption{Quartet of MASH-II PN, PN G322.4-00.1a (MPA\,1523-5710).
The nebula is bipolar in H$\alpha$
(greyscale image).  At short wavelengths (3.6\,$\mu$m [top left]) and
(4.5\,$\mu$m [bottom
left]) the central star dominates.  This star, and/or the nebula centre,
brightens at (5.8\,$\mu$m [top right]) and
(8.0\,$\mu$m [bottom right]. Diffuse 8.0-$\mu$m emission is also
detected along the major axis of the nebula \label{mash2qtet}}
\end{figure}

\twocolumn
\noindent
lacked images at 
high resolution. However, the importance of   having a well-defined 
set of PNe believed to be bipolar, to contrast with  round nebulae, 
was motivation for us to re-examine all available CCD  images for known
PNe via the catalogues of PN images by Manchado et al.  (1996)  and 
Balick (2008,2009) in order to 
characterize them.  Table~\ref{knattrib} offers a
characterization of these nebulae using the identical scheme as for MASH
nebulae. For starlike, unresolved PNe, we indicate
simply an ``S'' as no morphological type can be 
extracted.  For resolved SEC nebulae the sizes were taken preferentially 
from Tylenda et al.  (2003) who measured them at the10\,percent isophotal level,
or from Ruffle et al.  (2004). When necessary, we measured SEC nebular
diameters to the 10\,percent level ourselves,
comparing the SHS (Parker et al.  2005), H$\alpha$ and Short Red sizes.
In the case of very small or optically faint PNe we took the
IRAC 8.0-$\mu$m dimensions.

Both MASH-I and MASH-II PNe are very similar in their MIR appearance.
For example, in  Fig.~\ref{mash2qtet} we present a quartet of IRAC
images for MASH-II bipolar object MPA1523-5710 (PN G322.4-00.1)
overlaid on the greyscale SHS  H$\alpha$ image of this nebula.   In
Paper-I we illustrated several examples of such quartets to show the
changing morphology with IRAC wavelength.  The central star dominates
the nebular core at short wavelengths, brightening and growing in size 
at longer wavelengths as do many PNe that show the PAH emission 
features.  Note that  diffuse emission is also detected at 
8.0\,$\mu$m, along the major axis of the nebula.

\section{PNe of large angular size}
Both MASH-I and MASH-II include significant numbers  of PNe with large 
angular size29, but these are predominantly of much lower surface brightness 
than pre-MASH nebulae.  Large  angular size PNe are well-resolved
so that we can trace the variations in dominant MIR emission across 
the nebulae by MIR false colour changes. These PNe are of low surface 
brightness  but are close to the sun and so appear with large angular extent. 
They typically lack PAH emission, 
perhaps because these lower density nebulae are optically thin to ionizing 
radiation and their nuclei no longer furnish enough UV radiation to 
maintain easily detectable PDRs.
Although IRAC surveys of PNe (e.g. Hora et al.  2004) do readily detect 
nearby, high surface brightness optical PNe of large angular size, we 
found a dearth of MIR detections of large nebulae in Paper-I . We 
concluded  that the large angular MASH-I PNe in which PAHs
have been detected are bipolar, high-excitation PNe, 
in which PAHs are found in a high-density central circumstellar disk. 
We emphasize the fact that all large nebulae do not necessarily present the 
same MIR emission process, and offer two examples to demonstrate this.

\subsection{Abell 48} 
\begin{figure}
\vspace{8cm}     

\includegraphics{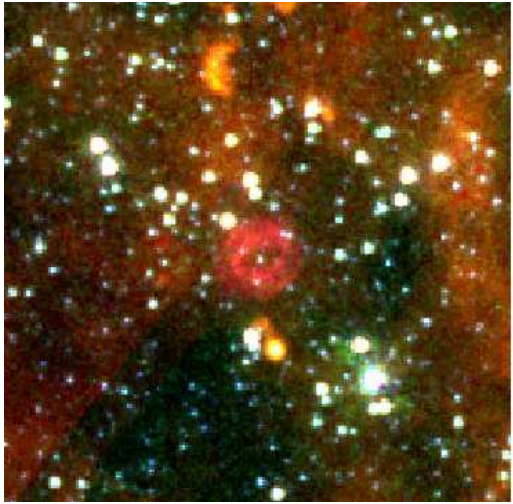} 
\caption{PN G029.0+00.4 (Abell 48) has a 50\,arcsec diameter in H$\alpha$ 
and MIR emission  is clearly seen across  the entire nebular 
extent. this PN is very clearly distinguished from the surrounding 
ISM by both false colour and shape.\label{abe48}}
\end{figure}

Fig.~\ref{abe48} shows the 50$^{\prime\prime}$ diameter PN G029.0+00.4 
(Abell 48), almost all of whose interior is red in false colour. 
This image also serves as 
a paradigm for a PN that stands out against the local  ISM.  There are red
regions elsewhere in Fig.~\ref{abe48} but these are clearly amorphous 
(e.g. the NW corner and W side of the box) or local diffuse orange
 patches (e.g. the N edge  of the
frame.  But this PN contrasts with everything else in the overall image 
by its red color which traces the entire extent of this PN. It is also 
of interest because it seems to be a member of the group of PNe with Wolf-Rayet 
central stars, (DePew et al. 2010), perhaps even of the rarer 
subset of [WC/WN] PNe (Morgan, Parker \& Cohen (2003); Todt et al.  (2010). 
Then one would expect the red nebular false colour to be PAH emission rather than H$_2$
because the C-rich aspect of  dual-dust PNe is shown by strong PAH bands. 
 
\subsection{Hen\,2-111} 
\begin{figure}
\vspace{8cm} 
\includegraphics{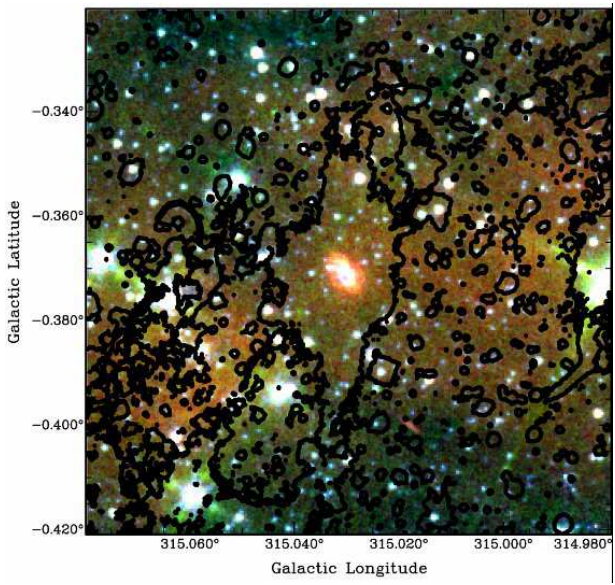} 
\caption{10$^{\prime}$ by 10$^{\prime}$ field containing the false 
colour image of PN G315.0-00.3  (Hen\,2-111).  This PN has an optical geometric diameter 
of over 10\,arcmin. MIR thermal dust 
emission is likely to cause the white region at the centre of this PN.
No other MIR emission appears to be associated with it.
Black contours show the extensive H$\alpha$ filaments. \label{bigpncore}}

\end{figure}

By contrast, PN G315.0-00.3 (Hen\,2-111: Webster et al.  1978) consists
of a bright central body (29$\times$15\,arcsec), surrounded by a huge 
bipolar filamentary halo of emission over 10\,arcmin in extent that 
is expanding at substantial velocities of
order 500\,km\,s$^{-1}$ (Meaburn \& Walsh (1989).  While these are all 
unusual characteristics for  a PN, this object is
still regarded as a bona-fide PN.  MIR 
emission from heated dust (false  colour white) is seen to be associated 
solely with the bright core (Fig.~\ref{bigpncore}).  The object 
lies in a region of the ISM with very complex structure but no other 
obviously associated MIR features appear within the optical extent of the nebula.  
The extensive network of H$\alpha$ filaments in the outer halo 
is represented by the black overlaid contours.

PN G315.0-00.3 offers an opportunity to compare the MIR colours of a 
bright PN core with those of the whole nebula.
The six IRAC colours (as presented in Table~\ref{derived}) for the entire 
nebula are (1.36,1.74, 2.37, 0.38, 1.02, 0.63)
but the bright compact core has (2.20, 3.49, 5.82, 1.29,  3.62, 2.34), 
presenting a substantially different MIR character. This implies sizeable 
spatial variations in the mix of emission processes across the PN.
Such processes will include PAH molecular bands, H$_2$ emission lines, 
fine structure lines, atomic recombination lines, thermal emission from dust
species, such as silicates, both amorphous and crystalline, carbon, and 
SiC, and species without any obvious MIR features, such as graphite.
On the basis of longer wavelength comparisons of Spitzer 24-$\mu$m 
and H$\alpha$ images of PNe, Chu et al. (2009) argued that the 
relative contributions of fine structure lines and dust emission vary 
from nebula to nebula. Hora et al. (2004) have observed that, even 
within a single PN, different MIR emission processes occur in 
different spatial regions.

\section{Analysis of the clean sample of PNe}
Table~\ref{derived} contains all our derived and collateral data on each
PN that has passed our careful selection criteria.  This information
consists of the
following: after the designation of a PN (col.1) we give the [3.6], theIRAC 
3.6-$\mu$ magnitude, the IRAC colour
indices in Vega-based magnitudes (cols.3-8).
col(9) contains the IRAC 4.5-5.8-8.0\,$\mu$m (blue, green, red) false
colour if the nebula stands out against the background.  
Cols.(10-11) present two optical emission-line
flux ratios: log$_{10}$H$\alpha$/[S{\sc ii}] (col.10) and 
log$_{10}$H$\alpha$/[N{\sc ii}] (col.11); the
electron density N$_e$ derived from the [S{\sc ii}] optical emission line
ratios  from the available spectroscopy appears in (col.12) ; (note we have 100\,percent
spectral coverage for all MASH PNe); whether a PN has radio  continuum emission ``m" from
Molonglo  at 843\,MHz  or ``n" from the NVSS at 1.4\,GHz) is noted in
(col.13); if so then the radio flux density is given in (col.14);  the
8.0-$\mu$m  IRAC flux density  appears in (col.15); if both MIR and radio
detections are available then the ratio of MIR/radio flux densities is
noted in (col.16). Our optical emission-line database was assembled from homogeneous
observations with the Flames spectrograph at the VLT, the 2DF and AAOmega 
instruments on the 3.9m Anglo-Australian Telescope, the Double Beam 
Spectrograph at the ANU 2.3m telescope, FLAIR on the UK Schmidt Telescope, 
and the grating spectrograph on the 1.9m SAAO telescope. For details see 
Parker et al.  (2006),  Miszalski et al.  (2008b, 2009a).

\subsection{False-colour {\it Spitzer} imagery} 
In our enlarged, clean  sample of 136 PNe, 60\,percent (82/136) stand out
against the ISM background in the IRAC images by their MIR false colours,
as described in \S2.1.  16\,percent of our PNe (22/136) lack the IRAC data
needed in all 3 bands required to create a false colour image while
23\,percent  (31/136) have IRAC data but cannot be separated from the diffuse
ISM emission in the MIR. We designate
these false colours by  O, R, and V to denote the three dominant false colours that we
have noted: orange, red, and violet, respectively when we combine the 
IRAC bands to create color images.  By orange we mean a mix of 5.8 and
8.0\,$\mu$m; red, dominantly seen at 8.0\,$\mu$m; and violet, a mix of
3.6 and 8.0\,$\mu$m.   A  ``u"  in the false colour column of Table~\ref{derived}
signifies that a PN lacks some of the IRAC data required to create the
false colour image. Finally, we use an ``X" in this column to indicate that,
despite having all the requisite IRAC data (i.e. detections at 4.5, 5.8
and 8.0\,$\mu$m)  there isinsufficient false colour contrast for a PN 
to be recognized in the MIR against the local background.

\subsection{Measuring the IRAC colours of PNe}
Techniques identical to those developed in Paper-I  were used to assess
the local sky brightness around each PN.  We overlaid  the IRAC images
 outside the contours of the H$\alpha$ image to accurately represent the
nebular area, and removed nearby GLIMPSE point sources to provide
cleaner sky subtraction.  This latter method works only
when the central region of the PN is not itself pointlike
and there is no obvious MIR central star.  Otherwise the 
diffuse photometry of the residual PN image significantly underestimates 
the nebular flux because the core will have been
removed from the image.  The sole difference in approach 
compared with Paper-I was that we integrated the MIR flux over the 
area of a PN guided by the H-alpha image even when visual inspection 
failed to identify the object in an IRAC image. The lower bound on 
nebular MIR flux is set by the outer boundary of the H-alpha PN 
image.  The outer bound is where the PN is no longer detected 
above the sky background in the {\it MIR}.

Estimates of the extended source aperture corrections required to
mitigate the effects of scattered light within IRAC on diffuse photometry 
have changed over the years since the launch of Spitzer.
 We have updated the diffuse region colours of Paper-I to the latest
(Nov.8, 2007) extended source calibration factors for IRAC, documented 
by Jarrett \footnote{http://spider.ipac.caltech.edu/staff/jarrett/irac/calibration/index.html} so that all our diffuse photometry is consistent with a common set of
current best practices.
For some nebulae we were able to compare our MIR magnitudes with
measurements of other groups, for example,
Hora et al.  (2004), Kwok \& Zhang (2008), or Phillips \& Ramos-Larios
(2009). Direct comparisons
based on the latest factors were not always possible.  Phillips \&
Ramos-Larios (2009) still used correction
factors from Reach et al.  (2005), which introduce magnitude offsets of
up to 0.1m compared with the usage of
current corrections. Kwok \& Zhang (2008) make no mention of  which
aperture corrections they applied to
their MIR photometry, potentially leading to differences as large as
0.33~mag (at 8.0$\mu$m), if no corrections were made.
There appears to be accord between the various data
sets subject to these uncertainties. Those
objects whose photometry shows differences that are not attributable to
aperture corrections are large nebulae. Here we have integrated over the 
entire nebula using the overlaid H-alpha image to set the inner edge of 
the PDR and integrating outwards until the PN is no longer detected 
above the local MIR sky background.  \emph Note that, for PNe of 
large angular size, others have measured only the MIR bright central 
region.  Hen\,2-111 (PN G315.0-00.3)  (\S5), 
is a good example.  Kwok \& Zhang
(2008:  cited by Hora et al.  2008a) measure an extent only
10$-$15$^{\prime\prime}$ for this object in the MIR.  The bright nebular
core in H$\alpha$ is 29$^{\prime\prime}$$\times$15$^{\prime\prime}$ but
the PN is seen in deep SHS  H$\alpha$ imaging to have a size of order
600$^{\prime\prime}$.  The associated MIR emission of the entire
nebular is 2 to 3\,mag brighter than the bright, dust-dominated core.
A less extreme example that indicates that MIR emission in an PN should
be sought over at least the area in which H$\alpha$ is apparent 
is Hen\,2-84 (PN G300.4-00.9). The H$\alpha$ size is 
36$^{\prime\prime}$$\times$24$^{\prime\prime}$ and the IRAC emission 
corresponding to the H$\alpha$ area is fully 1\,mag brighter in all 
four bands than that measured in the 10$^{\prime\prime}$ region noted
by Kwok \& Zhang (2008).

The IRAC colours in Table~\ref{colours} are calculated from the
differences between the diffusely
calibrated photometric magnitudes for a given PN.  If a PN was detected
in only one IRAC band then no
colours can be derived; e.g. MASH-I PN, PHR1635$-$4654 was detected
only at 5.8\,$\mu$m.
The quantities tabulated for each subset of PNe are the 6 median IRAC
colours,
each with its standard error of the median (sem), the number of PNe
contributing a
colour to each median, and the total number of nebulae that belong to
each subset.  Individual PN IRAC magnitude uncertainties are 
between 0.02m and 0.15m in bands 1,2 and between 0.07m and 0.23m 
in bands 3,4. These depend primarily on the local variations of the 
surface density of stars in IRAC bands 1 and 2, and on sky background 
variations in bands 3 and 4. These errors include the uncertainties in 
the IRAC aperture corrections applied to both sky and PN photometry. 
The resulting errors in our colours could be as little as 0.03 for $[3.6]-[4.5]$ 
to as large as 0.24 for $[5.8]-[8.0]$.

Table~\ref{colours} presents 17 different subsets of the PN sample that we
have constructed
according to various PN properties we are investigating, together with
all PNe combined. The colours of our MASH-I
and MASH-II nebulae, and of all MASH PNe (MASH-I and MASH-II) are
compared with those of previously
known nebulae.  Other subsets represented are those PNe that appear
orange, red, or
violet in their IRAC false colours; all PNe contrasted against the ISM
having any of these three false colours.  These can be compared with the colours of
all PNe that are detected in the MIR but are not distinguished from the ISM by visual
inspection of the images.

\subsection{The IRAC colours of different PN subsets}
Comparing MASH-I and MASH-II nebular colours  from
Table~\ref{colours} it is clear that five of the six IRAC colour indices
are apparently indistinguishable between these two MASH samples.
The remaining colour, [4.5$-$5.8], differs marginally  (3.3\,$\sigma$)
between MASH-II PNe and other sets.  The sense of 
this difference is that  [5.8] is
brighter or that  [4.5] is fainter in MASH-II PNe.
\S6.2.1 shows that our MASH-II PNe sample has a far smaller fraction of
bipolar nebulae (13\,percent) than
both the MASH-I sample (29\,percent) and the SEC set (34\,percent). Note that this may
be an underestimate of the true
MASH-II PNe bipolar fraction as there are a higher fraction of compact
objects  in MASH-II that has made morphological classification difficult.
Nevertheless, the typical [4.5$-$5.8] colour for the general MASH-II PNe
sample is statistically the same as that for all non-bipolar nebulae in our
combined sample (they differ by only 1.6\,$\sigma$) so that the
bipolar nebulae that are present are not sufficiently numerous to affect
this colour balance.

Ybarra \& Lada (2009) note that, in dissociative (J-type) shocks
that heat gas to above 4000\,K, CO is vibrationally excited,
as are [Fe{\sc ii}]  fine structure lines.  Both the CO emission lines
and most of the
[Fe{\sc ii}] lines within the overall wavelength range of IRAC fall
into the IRAC 4.5-$\mu$m bandpass. 
While we understand the difference between the discovery space for
MASH-I and MASH-II PNe, one
might  wonder whether there are also spectral differences in the
dominant  MIR emission mechanisms
of PNe in these samples.  In the NIR, most PNe spectra are dominated by
hydrogen recombination lines (Hora et al.  1999, their Table 1) while the
next most frequent set of PNe show both hydrogen recombination lines and
those of H$_2$.  These authors list very few nebulae that are dominated
by H$_2$ lines and these tend to be unusual or rare objects, such as
edge-on bipolars or proto-planetary nebulae.
PNe in the MIR reveal more fine structure lines and strong, broad, dust
emission features are common so it is
highly likely that most PNe, depending on their evolutionary state,
inherent metallicity and morphology,  will display a variety of MIR
emission processes, rather than being dominated by a single mechanism.

Among the IRAC false colour subsets for the PNe samples we have created
we see only two marginally significant (3.0$\sigma$) differences.
PNe with a predominant orange false colour differ statistically from
violet PNe only in the [3.6$-$8.0] colour.  Because the distinction
between orange and red false colours is somewhat subjective, we decided
to merge the orange and red looking PNe and compared their combined colours
with those of violet PNe, which presents the same difference in [3.6$-$8.0].
A broader version of this search for colour differences 
was executed by seeking colour distinctions between the subset of
82 PNe that are visually contrasted against their surroundings and the
31 PNe that reveal no such contrast.  Here too there is a  single
distinction (3.0\,$\sigma$) in that those PNe that stand out against
the ISM have a smaller median [4.5$-$5.8] colour in the same sense as
MASH-II PNe compared with other PN samples  (discussed above).

\subsubsection{IRAC colours of the general PN population}
Stanghellini et al.  (2000) found for LMC PNe that morphology is a
good indicator of the progenitor population.  Therefore we carried out
a related test, namely to look for IRAC colour differences between the
samples of Type~I and non-Type~I PNe that we have determined based on our
comprehensive optical spectroscopy (see  Frew \& Parker (2010; their Fig.4). 
It is already known that there appears to be a strong correlation 
between a PN being Type~I and also having  bipolar morphology.  
However, for these comparisons too, no meaningful  colour differences were found.
 Another aspect of Type~I PNe in our sample that is of interest is the 
proportion of these nebulae to be found in different regions.   A
volume-limited empirical estimate from PNe within 1\,kpc of the sun gives 10\,percent
(Frew, Parker \& Russeil 2006), while Moe \& De Marco (2006b) estimated theoretically
a value of 6$\pm$4\,percent, assuming that stars above 4M$_\odot$ are the progenitors 
of Type~I PNe.  We find 25\,percent of our clean sample are  Type~I PNe.  Therefore,
GLIMPSE confirms that a Galactic latitude-selected sample of PNe 
is dominated by high-mass stellar progenitors.

The largest morphological group of nebulae is represented  by elliptical
PNe that constitute 48\,percent of the clean sample (65/136).
There are only 14 round, 5 irregular, and 4 asymmetric PNe in the clean
sample.  But the ensembles of round and elliptical PNe also display 
no colour distinctions.  The small group of PNe for which high spatial 
resolution imagery is lacking are not distinguished from any others 
by their colours either. Beyond the few distinctions cited above we 
found no additional significant colour differences in this table 
between  groups of PNe.

\subsubsection{Bipolar versus non-bipolar PNe}
Of particular interest are the colours of bipolar versus non-bipolar PNe.
Therefore, we have separated all bipolar nebulae from 
non-bipolar nebulae  to determine whether this aspect of morphology 
is reflected in the MIR colours of PNe. 
In our clean sample of MASH-I nebulae, 29\,percent appear bipolar (19/65). For
MASH-II PNe the corresponding number is 13\,percent (4/31).  Of our MASH-I 
and MASH -II nebulae combined,  24\,percent are bipolar (23/95).  This rises
to 34\,percent (15/41) for the SEC objects, probably because these earlier
discoveries are brighter nebulae whose morphologies were more 
readily recognized. Overall the total GLIMPSE-selected  clean PN 
sample contains  27\,percent (37/136) bipolar PNe.
For comparison we have assembled a non-bipolar subset by excluding all PNe
with a ``B'' descriptor
and all those characterized as ``S'' (i.e. unresolved or point-like
nebulae) of which there are 13 instances.  Morphologies for these
compact objects remain indeterminate.  There are thus 87 non-bipolar
objects, accounting for 64\,percent of the sample.  

Bipolar PNe contain dusty disks.  Some of these are likely to present 
an edge-on dusty disk aspect in our direction. In crudest terms this
suggests the prospect of a difference in the range of dust temperatures
viewed and, therefore, revealed in the MIR.  Edge-on dusty
disks might appear cooler and redder than PNe lacking either flattened
structures or with strongly inclined disks.  Based on Gatley's Rule 
(\S4), which states that only bipolar PNe
show strong H$_2$ emission in
the near-infrared, one might have anticipated an analogue for the IRAC
4.5-$\mu$m band which encapsulates such features. Somewhat surprisingly,
we see no such difference.  However, the rule was
based on imaging in the 2.122\,$\mu$m line and the inference is that in
the equatorial dusty disks of PNe
molecular material ejected by red giant precursors survives and is
subsequently shocked by ionized outflows
(e.g. Kastner et al.  1996).  Ybarra \& Lada (2009) have computed the
strengths of H$_2$ lines that arise in low
temperature shocks, where H$_2$ is the dominant coolant, and that fall
within the IRAC bands.   Whether the
emission arises through shocks or by  fluorescence one expects all the
IRAC bands to contain H$_2$ lines.
Indeed, Hora et al. (2004) and Hora \& Latter (1996)  have
commented on the presence in PNe of the
 S(9) line at 4.69, the S(7) and  S(5) lines  at 5.51 and 6.91\,$\mu$m,
and two additional strong
lines in the 8.0-$\mu$m band.  Consequently, the multiple contributions
to the IRAC bands from these various H$_2$ emission lines make it
unlikely that H$_2$ would dominate any specific IRAC colour signature
for any PN ensemble.   Based on our 136 PNe, there  is no unique 
MIR colour that serves as an indicator
of bipolar PNe, like Gatley's Rule for the presence of H$_2$ in the NIR.
However, Miszalski et al.  (2009b) note a tendency for bipolar PNe to
be associated with 8.0-$\mu$m emission in their waists but no other 
prominent IRAC emission.  The low surface brightness of MASH PNe
might militate against IRAC detections of such nebular waists but  
H$_2$ might be bright enough to be detected in some PNe. A possible
analogue might be RCW69 (Paper-I; Fig.12) which is a large faint PN
associated only with weak IRAC emission at 8.0\,$\mu$m, albeit slightly displaced
from the PN waist.

Consequently, one firm result from our study is that there do not
appear to be any clear MIR distinctions between the IRAC colours 
of the overall PNe population, whether segregated by discovery 
method, MIR false colour, or optical morphology.  
But PN contaminants do exhibit such distinctions (\S6.4).
Therefore, the general MIR colours  of the combined MASH and previously
known PNe sample appear to be highly robust and representative 
of the PN phenomenon.

\begin{table*}
\caption{Diffusely calibrated median colours (in Vega-based magnitudes)
of PNe in
the six IRAC indices for different subsets of PNe and for all nebulae
combined.
Median and standard error of the median (sem) are given together with
the number of PNe that contribute
to a colour and the total number of PNe available in that
sample. \label{colours} }
\begin{tabular}{cccccccc}
Subset&     $[3.6]-[4.5]$& $[3.6]-[5.8]$& $[3.6]-[8.0]$& $[4.5]-[5.8]$& $[4.5]-[8.0]$&   $[5.8]-[8.0]$&  Sample   \\
      &           No.&           No.&           No.&  No.&           No.&             No.&      No.\\
\hline
All PNe& 0.81$\pm$0.08&  1.73$\pm$0.10&  3.70$\pm$0.11&  0.86$\pm$0.10&
2.56$\pm$0.11&  1.86$\pm$0.07\\
        &    107&         110&         106&         115&
121&         116&    136\\
MASH-I& 0.63$\pm$0.15&  1.88$\pm$0.22&  3.33$\pm$0.23&  1.20$\pm$0.17&
2.69$\pm$0.20&  1.74$\pm$0.14\\
       &          41&          42&          40&          49&
55&          50&     65\\
MASH-II& 0.78$\pm$0.12&  1.43$\pm$0.10&   3.19$\pm$0.16&
0.48$\pm$0.14&  2.33$\pm$0.13&  1.88$\pm$0.08\\     \
      &          28&          28&          28&          27&
28&          28&     30\\
MASH&  0.70$\pm$0.10&  1.72$\pm$0.14&  3.22$\pm$0.15&  0.96$\pm$0.12&
2.45$\pm$0.14&  1.81$\pm$0.10\\
     &          69&          70&          68&          76&
83&          78&     95\\
Known&  0.92$\pm$0.13&  1.73$\pm$0.14&  3.86$\pm$0.15&  0.83$\pm$0.16&
3.07$\pm$0.16&  2.11$\pm$0.09\\
      &          38&          40&          38&          39&
38&          38&     41\\
Orange& 0.72$\pm$0.11&  1.84$\pm$0.10&  3.81$\pm$0.14&  1.07$\pm$0.15&
3.11$\pm$0.17&  1.97$\pm$0.09\\
      &          26&          26&          26&          27&
27&          27&     27\\
Red& 0.92$\pm$0.09&  1.49$\pm$0.13 & 3.69$\pm$0.14&  0.64$\pm$0.13&
2.45$\pm$0.12&  2.02$\pm$0.08\\
       &       38&          38&          38&          37&
38&          38&       38\\
Violet& 0.51$\pm$0.14&  1.30$\pm$0.26&  2.84$\pm$0.29&  0.63$\pm$0.40&
1.86$\pm$0.41&  1.71$\pm$0.16\\
      &          15&          15&          16&          16&
17&          17&     17\\
O$+$R&   0.81$\pm$0.09&  1.71$\pm$0.09&  3.75$\pm$0.10&  0.78$\pm$0.10&
2.69$\pm$0.10&  1.99$\pm$0.06\\
     &          64&          64&          64&          64&
65&          65&     65\\
Contrast& 0.81$\pm$0.06&  1.70$\pm$0.09&  3.69$\pm$0.10&
0.74$\pm$0.11&  2.46$\pm$0.12&  1.86$\pm$0.06\\
      &        79&             79&            80&          80&
82&           82&       82\\
No contrast& 0.73$\pm$0.28&  1.95$\pm$0.31&  3.81$\pm$0.36&
1.38$\pm$0.18&  3.07$\pm$0.28&  1.86$\pm$0.14\\
             &          24&          25&          24&    31&          31&          31&     31\\
Type~I&    0.81$\pm$0.21&  1.71$\pm$0.24&  3.21$\pm$0.26& 0.97$\pm$0.22&  2.77$\pm$0.28&  1.68$\pm$0.16\\
             & 27&          28&          28&          33&  33&          32& 34\\
Non-Type~I& 0.80$\pm$0.08&  1.74$\pm$0.11&  3.75$\pm$0.12& 0.85$\pm$0.10&  2.54$\pm$0.10&  1.91$\pm$0.08\\
             &      80&          82&          78&          82&  88&          84&  102\\
Bipolar& 0.98$\pm$0.15&  2.04$\pm$0.21&  3.68+0.22&   0.97$\pm$0.23& 2.51$\pm$0.30&  1.68$\pm$0.19\\
      &          28&          30&          29&          31&    32&          33&     36\\
Non-Bipolar&  0.71$\pm$0.11&  1.52$\pm$0.12&  3.46$\pm$0.15& 0.76$\pm$0.11&  2.46$\pm$0.11&  1.86$\pm$0.08\\
       &          66&          67&          64&          71&   76&          70&     87\\
Elliptical& 0.87$\pm$0.13&  1.70$\pm$0.16&  3.75$\pm$0.17& 0.74$\pm$0.14&  2.67$\pm$0.13&  1.92$\pm$0.08\\
           &          51&          50&          49&  54&          56&          53&     65\\
Round& 0.68$\pm$0.23&  1.35$\pm$0.15&  3.09$\pm$0.22&  0.86$\pm$0.25&
2.22$\pm$0.31&  1.77$\pm$0.24\\
          &          11&          13&          11&          13&  13&          13&     14\\
Starlike& 0.79$\pm$0.07& 1.76$\pm$0.24&  3.87$\pm$0.25&  1.00$\pm$0.21&  3.12$\pm$0.23&  2.05$\pm$0.07\\
            &          13&          13&          13&  13&          13&          13&     13\\
\hline
\end{tabular}
\end{table*}

\begin{table*}
\caption{Diffusely calibrated median colours (in Vega-based magnitudes)
for H{\sc ii} region contaminants, whether ultra-compact, compact or diffuse.  
Median and standard error of the median(sem) are given. \label{hii}}
\begin{tabular}{ccccccc}
Subset&     $[3.6]-[4.5]$& $[3.6]-[5.8]$& $[3.6]-[8.0]$& $[4.5]-[5.8]$&
$[4.5]-[8.0]$&   $[5.8]-[8.0]$ \\
\hline
UCHII{\sc ii}&        1.49$\pm$0.12&  2.50$\pm$0.12&  3.57$\pm$0.12&
1.01$\pm$0.12&  2.08$\pm$0.12&  1.07$\pm$0.12 \\
 Diffuse \& compact H{\sc ii} &  0.47$\pm$0.06&  3.08$\pm$0.10&  4.85$\pm$0.10&
2.60$\pm$0.07&  4.43$\pm$0.05&  1.84$\pm$0.05 \\
\hline
\end{tabular}
\end{table*}

\subsection{Colour-colour planes}
In Paper-I we illustrated that the nominal box occupied by PNe in MIR
colours suffered very little contamination by other types of point source,
 although the disposition of diffuse  and compact H{\sc ii} regions in at least one IRAC
colour-colour plane might encroach on the predicted location.  We have
now explored this important issue in some detail as it affects our
ability to provide robust loci for true PN in selected colour-colour
planes.
One problem area is for unresolved or barely resolved PNe which could be
contaminated by a wide range of common Galactic point source emitters.
We must also  explore the isolation of IRAC colours of resolved PNe from
the various kinds of possible diffuse contaminants.  The IRAC colours
are available empirically for  diffuse and compact H{\sc ii} regions
(Cohen et al.  2007b) while synthetic colours can be estimated for
UCHIIs  from spectral libraries such as
that inherent in the SKY model of the point source sky (Cohen 1993).  
Table~\ref{hii} gives both these colour sets.
We then examined the formal metric separations between the median
colours of our entire PN sample (Table~\ref{colours}: ``All PNe") and
those of  diffuse and compact, and UCHIIs, seeking the statistically
most significant colour differences.  This analysis yields the optimal
MIR colour-colour planes for separating PNe from these common types of
interloper.

This approach is justified by calculations of the varying surface density 
of contaminants throughout the plane at latitude zero in the GLIMPSE region.
At 8.0\,$\mu$m, to IRAC's local confusion-limited mag, 
SKY indicates  that UCHIIs contribute 
from 20 sources deg$^{-2}$ at longitude 10$^{\circ}$, 10deg$^{-2}$ 
at 330$^{\circ}$, to 2deg$^{-2}$ at 65$^{\circ}$.  Diffuse H{\sc ii} 
regions would be in addition to these point sources.

Table~\ref{hii} contains two sets of colours for the objects most frequently confused 
with PNe, namely  H{\sc ii} regions.  The first line offers the synthesized
median colours (in Vega-based magnitudes) of UCHIIs 
based on empirical templates for these objects.  The 
synthetic photometry was carried out using the newest relative spectral 
response curves for the IRAC bands (see Hora et al.  2008b). 
The second line presents the diffusely calibrated colours of  diffuse and 
compact H{\sc ii} regions (whose colours are indistinguishable).  These data were 
used to plot the corresponding boxes in our colour-colour planes that 
represent the median locations of these contaminants in the three planes which we discuss below.  

Table~\ref{hii} shows that PNe are well separated from UCHIIs, their 
chief contaminants, in $[3.6]-[4.5]$, $[3.6]-[5.8]$ and $[5.8]-[8.0]$ by 5, 5 and 6\,$\sigma$, 
respectively, and marginally (3\,$\sigma$) in $[4.5]-[8.0]$.  The 
metric distances between the median colours of PNe  and diffuse H{\sc ii} 
regions are substantial in all their colours except $[3.6]-[4.5]$, which is just over 3\,$\sigma$.
The differences  in $[3.6]-[5.8]$, $[3.6]-[8.0]$, $[4.5]-[5.8]$, $[4.5]-[8.0]$ 
and $[5.8]-[8.0]$ are 5, 10, 8, 14 and 15\,$\sigma$, respectively.  
These conclusions guide the choice of the optimal colour-colour planes for the 
recognition of PNe.

 We attribute these
different degrees of colour distinction between the two kinds of H{\sc ii} 
region and PNe to two aspects of their IR spectra.  First, deep 10-$\mu$m 
silicate absorptions characterize UCHIIs but are absent in 
diffuse regions where the central star is no longer MIR-bright. Secondly, 
the ratio of the 6.2-$\mu$m PAH band to that at 7.7\,$\mu$m is greater 
in H{\sc ii} regions than in PNe (Cohen et al.  1989). These effects reduce
all colour indices with respect to the 8.0-$\mu$m band in UCHIIs 
and enlarge those with respect to the 5.8-$\mu$m band in diffuse regions

\begin{figure}
\vspace{8cm}
\includegraphics{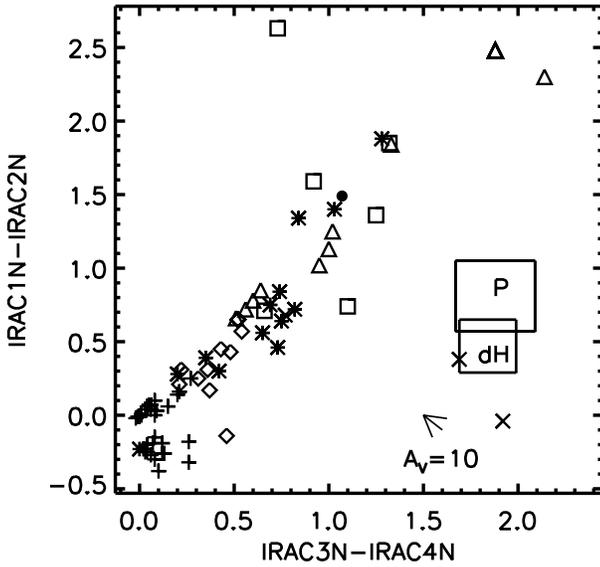}
\caption{Detail of the IRAC 1$-2$ vs. 3$-$4 plane comprising the
combined generic locations of 87 different types of IR point source. Key:
pluses - normal dwarfs, giants, supergiants; asterisks - AGB M stars;
diamonds - AGB
visible C stars; triangles - AGB deeply embedded IR C stars; squares -
hyperluminous
objects (these objects include deeply embedded OH/IR stars and
early-type hypergiants (Cohen 1993);
a small number are required to reproduce MIR source counts at low
latitude: Wainscoat et al. 
1992); crosses - exotica (T Tau stars, reflection nebulae); 
 filled circle - UCHIIs.
The reddening vector corresponding to an A$_V$ of 10\,mag is shown by
the shaft
of the arrow in the lower right corner. Two rectangles are also plotted
for the median\,$\pm$3\,sem
boxes for diffuse H{\sc ii} regions (labelled ``dH") and for the colours
of our entire PN sample
((labelled ``P"). The use of the ``N" in the IRAC band designations
(IRAC1N, etc) signifies the usage of
the latest relative spectral response curves for IRAC (Hora et al. 
2008b).\label{1234}}
\end{figure}

\begin{figure}
\vspace{8cm}  \includegraphics{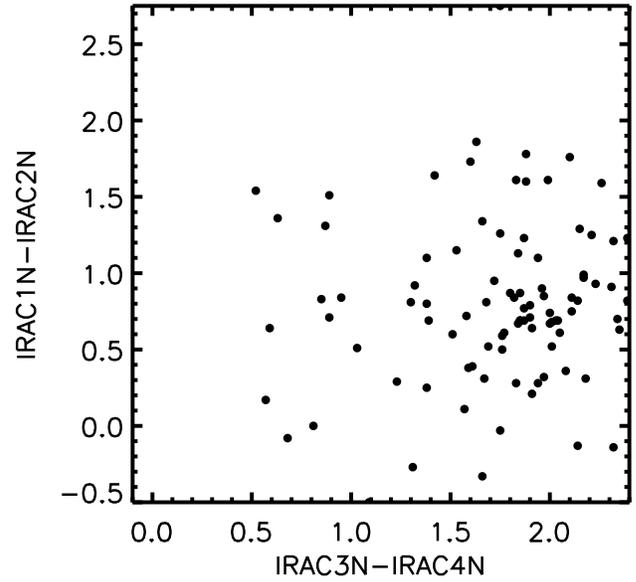}
\caption{As in Fig.~\ref{1234} but showing only the distribution of the 136 PNe in this plane.\label{1234full}}
\end{figure}

\begin{figure}
\vspace{8cm}  
\includegraphics{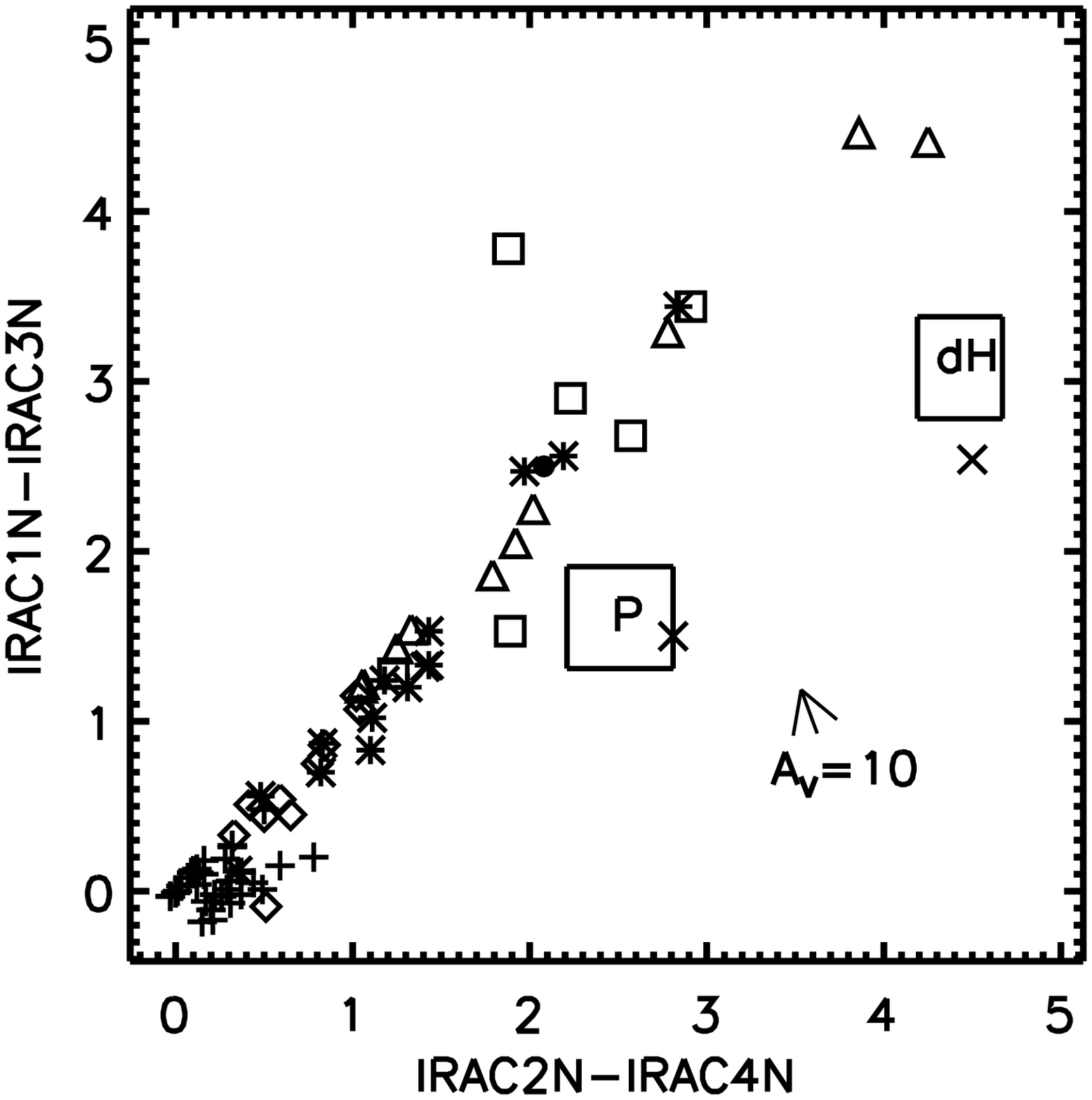}
\caption{As in Fig.~\ref{1234} but for the IRAC 1$-$3 vs. 2$-$4 plane.
\label{1324}.}
\end{figure}

\begin{figure}
\vspace{8cm}
\includegraphics{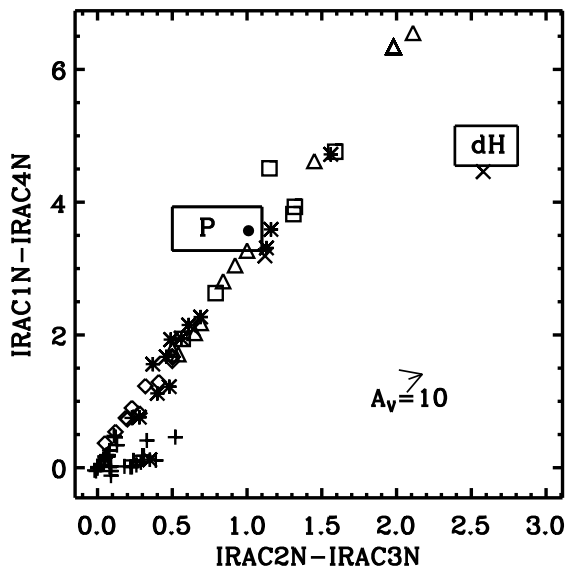}
\caption{As in Fig.~\ref{1234} but for the IRAC 1$-$4 vs. 2$-$3 plane.
\label{1423}}
\end{figure}

When performing our MIR characterisation of PNe we prefer to consider
PNe bright enough to be detected in all IRAC bands so that all  six
colour indices are available.  We avoid the degenerate planes in which
one IRAC band is common to the colour  indices on both axes, which can
lead to specious correlations.  We offer three different diagrams for 
IRAC colours:
[1-2] vs. [3-4], [1-3] vs. [2-4], and [1-4] vs. [2-3]
(Figs.~\ref{1234}, \ref{1324}, and \ref{1423}, respectively.  Each plot
also displays boxes which represent the median distribution of all PNe$
\pm$3\,sem  and similar boxes for diffuse H{\sc ii} regions (labeled`
P' and `H' in the diagrams), respectively.  
We also present Fig.~\ref{1234full}, showing our entire sample of 136 PNe in 
the Fig.~\ref{1234} plane to better present their overall distribution.

Although it is traditional to
plot PNe in the  [1-2] vs. [3-4] plane (e.g. Hora et al.  2004), one must
consider both the distance between UCHIIs
 (the  filled circle in all these colour-colour
plots) and the PN boxes, and that between diffuse
H{\sc ii} regions and the PN boxes.  The  [1-2] vs. [3-4] plot places the
PN box far from the  filled circle that
denotes the typical location of  bright unresolved H{\sc ii} regions but
there is clearly some modest overlap with diffuse H{\sc ii}
regions, as expected of these empirically most common contaminants.  For
well-resolved PNe the  [1-3] vs. [2-4]
plane separates their colours from those of diffuse H{\sc ii} regions.
There is some encroachment by reflection nebulae in this plane although
that is easily resolved by visual inspection of H$\alpha$ and Short Red continuum
images (a reflection nebula is about as bright in both images).
Likewise, for PNe that are
not of small apparent size, the [1-4] vs. [2-3]  plane most widely
separates the relevant objects but, ironically,
 the right-hand side of the PN box in this plane also includes the
location of UCHIIs.  Each of the three
colour-colour planes offers advantages and disadvantages. However, one
can readily select a suitable colour-colour plot that  would reduce the
level of a major contaminant. This is particularly true in the Galactic
plane, whether one is working with  unresolved or with large PNe. In
summary, IRAC MIR colour-colour planes can provide an effective means of
separating PNe from their primary contaminants, namely H{\sc ii} regions.

{\bf This paper is a proof of concept of a new approach to find
PNe using only IRAC colours. To test the method we retain
our strong rejection of potential contaminants. In future work
we shall hone the technique by balancing colour box extent
against rejection of non-PNe.}

\noindent
Those PNe that can be recognized against  the local ISM background
appear in only three MIR false colours when
using the IRAC 4.5, 5.8 and 8.0\,$\mu$m bands to encode blue, green, and
red, respectively.  These are distinct
from the false colours of H{\sc ii} regions in this same scheme,
offering a useful discriminant.  The three false
colours in these composite images are the same as were found in Paper-I.
Examination of the 82 PNe that reveal false colour above their surroundings 
shows the following frequencies of PNe by colour:
25 orange (32\,percent); 36 red (46\,percent; and 17 violet (22\,percent).  These 
proportions are very similar to those found for a sample of 47 LMC PNe.  These 
were observed with Spitzer's IRS (InfraRed Spectrometer) by Stanghellini et al.  (2007)
and by Bernard-Salas et al.  (2009).  We find that  42 of these are contrasted
against their local ISM.  Their false colours appear with similar frequency to
those of Galactic PNe.  The corresponding proportions for LMC PNe are  
32\,percent; 51\,percent; and 13\,percent.
That the same three false colours of PNe  occur in the LMC and in our 
Galaxy, and with rather similar frequencies, suggests that these 
PN colours arise anywhere from  the same mix of physical 
emission processes. 

\section{An optical emission line nebulae diagnostic diagram}
 SMB proposed a  diagnostic plane based on the ratios of 
common optical nebular emission lines as
a means to separate the various kinds of nebular objects according to 
their excitation mechanisms, physical sizes, and electron densities.
For example, Herbig-Haro objects and SNRs are shock-excited, while H{\sc 
ii} regions and PNe are photoionized and distinct domains
were identified  by SMB to show where SNRs and H{\sc ii} regions lie.  
The bulk of the area of this plane
was occupied by PNe whose range was bounded by two curves, both 
quasi-linear. Further,  PNe of larger physical scale were
located to the lower left of the region between the two lines, and 
smaller PNe to the upper right.  The PN region was extended
in both directions by dashed lines.  A subsequent rendition of the 
log-log plane of  H$\alpha$/[S{\sc ii} vs. H$\alpha$/[N{\sc ii}] became
known as the ``Canto diagram" (Canto 1981) but we shall cite this 
diagram as the SMB plot. The original SMB  PN domain was
empirically enlarged by Riesgo \& Lopez (2006) on the basis of a sample 
of 613 nebulae.  The purpose of this extension was
to apply this plot to modern PN catalogues that present a much larger 
range in PN line ratios than
was available to SMB. Riesgo \& Lopez (2006; their Fig.4) represent the 
extended range of PNe by an ellipse which accommodates
85\,percent of their PN sample.  However, this encroaches on almost half the 
area of the originally designated H{\sc ii} regions' box.  Therefore, we 
prefer
to use the quasi-linear bounds shown by Riesgo \& Lopez (2006; their 
Fig.1). Bipolar PNe are mostly located near the lower
bounding line.  Corradi et al.  (1997; their Fig.3) studied 19 such 
objects and presented the SMB plot.  58\,percent of these PNe fall below
the bounding line; the remainder are widely scattered throughout the plane.

In this paper we have used solely our own  optical spectra in 
generating equivalent diagnostic diagrams,
preferring to use a homogeneous data set with a uniform reduction 
process. For our purposes, it is convenient to use the
customary SMB diagram to highlight both similarities and differences
between groups of PNe.  However, a new and far more detailed version of 
this diagram based on a careful evaluation of about 3000
emission-line objects is being developed by Frew et al.  (2010, in prep.)
A preliminary version of this plot is presented by Frew \& Parker
(2010; their Fig.4), incorporating many new MASH PNe.
Those authors showed that Type~I PNe (following the KB94 definition)
inhabit a distinct region in the SMB plot, and are well separated
from non-Type I objects.  For those GLIMPSE PNe without abundance
determinations in the literature, we placed each PN onto our revised
SMB plot to determine if a nebula was Type I, based solely on the
relative de-reddenened strengths of the red emission lines in our spectra.

\begin{figure}
\vspace{8cm}
\includegraphics{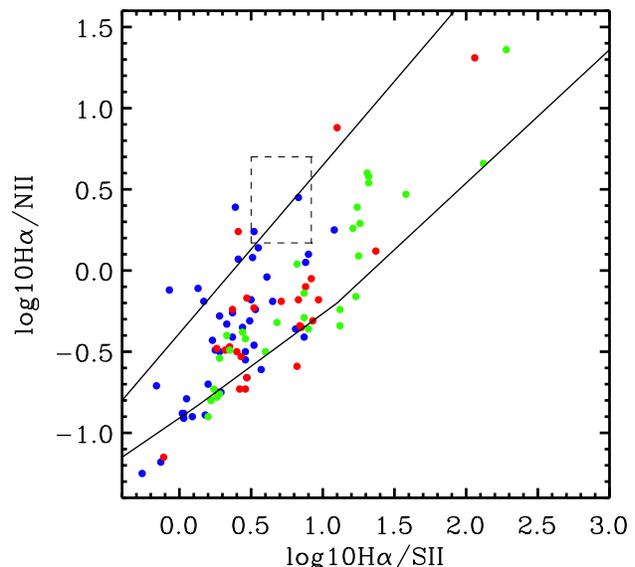}    
\caption{The SMB diagram for all PNe in our sample.  Symbols:
blue - MASH-I; red - MASH-II; green - previously known PNe. The dashed 
box is that defined by SMB for H{\sc ii} regions.  The two black lines are 
the bounds to the extended PN zone that accommodates the PN sample of 
Riesgo \& Lopez (2006).  \label{riesgo}}
\end{figure}

Our three samples of MIR detected PNe appear quite similarly distributed 
in the SMB  plot (Fig.~\ref{riesgo}).  Most lie between the
traditional bounding lines or slightly below the lower line.  The PNe 
known prior to MASH are rather more dispersed than
MASH nebulae, perhaps reflecting the heterogeneity and lesser accuracy 
of their line measurements compared with
the homogeneous values achieved by MASH follow-up spectroscopy.
For example, the two nebulae that lie in the upper right corner of the 
plane may  simply have very weak nitrogen lines,
elevating the ratio of H$\alpha$ to [N{\sc ii}].  We have omitted the 
region where SNRs are located but we note that
none of our nebulae lie within this zone.  A single object 
(PHR1544-5607) might lie on or close to the SNR boundary
($-$0.07,$-$0.12). It does have strong [S{\sc ii}] emission lines, is fairly 
indistinct in the optical imagery and is designated only as a possible 
PN in MASH.
We also indicate the commonly used dashed box that characterizes H{\sc 
ii} regions.  It is of interest that  
only two PNe are found inside this latter box, PHR1507-5925 and 
PHR1622-5038, perhaps validating
our efforts to eliminate H{\sc ii} regions from the PN sample.  The 
former is regarded as a True PN; the latter
has reasonable morphology for a PN, [O{\sc iii}], and no H$\beta$ in the 
blue so it is classified  as a Likely PN.

Had Mz~3 been plotted in Fig.~\ref{riesgo}
it would have appeared at (3.08,-0.09), far 
from all PNe,
confirming a further difference between its true character and those of 
accredited PNe.

\begin{figure}
\vspace{8cm}
\includegraphics{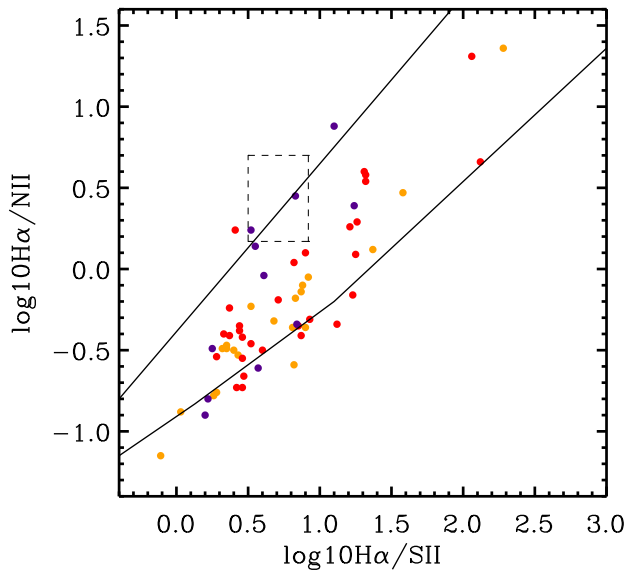}  
\caption{The SMB diagram for all PNe in our sample as for 
Fig.~\ref{riesgo} Symbols are colour-coded to match the red, 
orange, or violet MIR false colours of PNe. \label{riesgorov}}
\end{figure}
 
\begin{figure}
\vspace{8cm}
\includegraphics{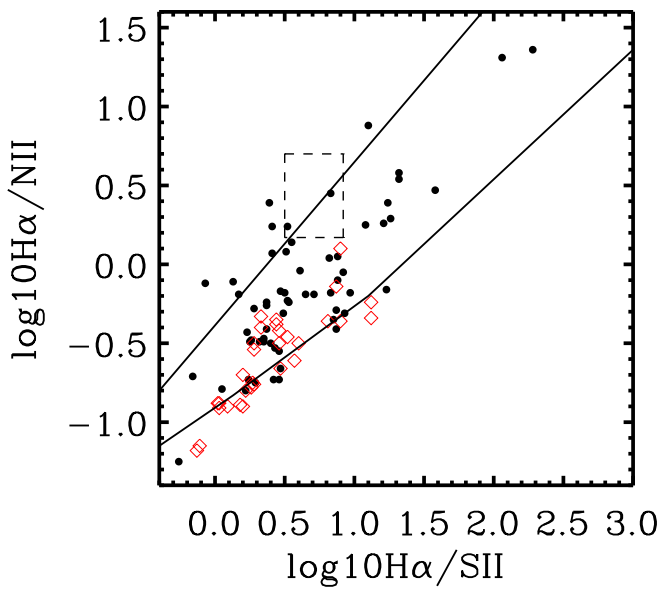}
\caption{The SMB diagram for all PNe in our sample as for 
Fig.~\ref{riesgo}  but distinguishing bipolar from non-bipolar nebulae. 
 Symbols: black - non-bipolar PNe; red - bipolar PNe.  \label{riesgobp}}
\end{figure}

Fig.~\ref{riesgorov} presents the same plane but now colour-coded 
according to the MIR false colour of PNe that
can be distinguished from their surroundings.  Red is the most commonly 
found false IRAC colour for PNe and these
are well-distributed throughout the PN zone.   By contrast, orange 
nebulae (with a single exception)
are located only from about the centre of the PN zone down to the lower 
left extreme of the PN zone. 
Violet false-coloured PNe are  more widely dispersed than orange PNe, 
despite their relative paucity.  
This  suggests that orange PNe are more evolved than other PNe. All 
three false colour samples appear to be equally likely to contain 
bipolar nebulae, based on their distribution  near and below the lower bounding line.

Our third variation on the SMB plot (Fig.~\ref{riesgobp}) distinguishes 
between bipolar (open red squares)
and non-bipolar (filled black circles) PNe. 
The distribution of bipolar PNe in our sample is much more restricted 
than that of the smaller Corradi et al.  (1997) sample.  We
find that bipolar nebulae occur only to the left of the centre of the PN 
zone.  The area of the entire PN zone that contains bipolar PN
is less that one third of the total area of the PN zone.  If our clean 
PN sample pertains to all PNe then this is a substantial
reduction in the area of the SMB diagnostic plot in which one might 
expect to identify PNe as bipolar.
Using our cleaned sample of PNe suggests that the diagnostic value of 
the SMB plane for bipolar PNe might be
sharper than previously thought,  perhaps due to the careful 
morphological classification of the MASH sample in particular,
even when based on the same imaging data.

\section{Physical radii of PNe}
We have derived intrinsic sizes (radii in parsecs)  for most of the PNe in our overall sample 
based on the calculated H$\alpha$ SB-distance relation (Frew, 2008;  Frew \& 
Parker 2010), using the observed angular dimensions, a measured H$\alpha$ flux and a reddening. 
Literature  distances, if available and deemed sufficiently reliable, are also used where necessary. 
Many of the H$\alpha$ fluxes and reddenings, (which will appear in a separate publication),  
although evaluated as carefully as current data will allow,
are preliminary, so that the distances (and hence radii) should currently 
be taken with some caution.  A few of the reddenings were determined from unpublished MASH
line fluxes, and others from a comparison of H$\alpha$  and radio fluxes.
To do this, we assumed that the 6-cm flux is the same as the flux density given in this paper.  

However, we hope that some  statistical conclusions might emerge from the 
ensemble of sizes when we employ PN radii as proxies for PN age.

We have chosen to assign our PN sample to a few selected size bins for
the present, although further efforts (Frew 2010, in prep.) will offer both 
new and improved determinations and refinements of the reddenings, 
fluxes, and sizes. The adopted colour code
to represent the bins is as follows: turquoise, very young PNe, radius $<$0.1\,pc;
blue, young, radius 0.1$-$0.2\,pc; green, middle-aged, 0.2$-$0.4\,pc; orange;
old PNe, 0.4$-$0.9\,pc; and red, senile PNe, $>$0.9\,pc.  We adopt a radius of
 0.9 pc rather than 0.8 pc to be consistent with the completeness radius of Moe \& De Marco (2006b).
Despite the input data being preliminary, we find clear trends in the SMB
diagrams with PNe evolving from upper right to lower left.

\begin{figure}
\vspace{8cm}
\includegraphics{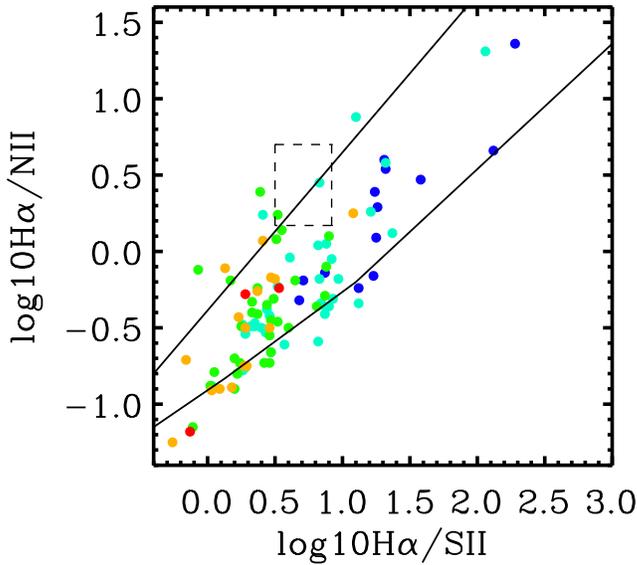}
\caption{The SMB diagram for all PNe in our sample for which we have assessed 
the distance and PN physical radius. Symbols: turquoise, very young; blue, young; green, 
middle-aged; orange, old; red, senile.  \label{riesgoage} } 
\end{figure}

\noindent
Table~\ref{agesize} indicates median PN radii for different groups of PNe.

\begin{table}
\caption{Median radii in pc for PNe as a function of age for each group. 
The median and standard error of the median (sem) are given.}
\label{agesize}

\begin{tabular}{cccc}
Subset&  No. of PNe&   median radius& sem\\
\hline 
Known   & 38& 0.12& 0.04\\
MASH-I  & 63& 0.34& 0.04\\     
MASH-II & 31& 0.15& 0.03\\ 
All MASH& 94& 0.25& 0.03\\ 
\hline
\end{tabular} 
\end{table}
 
Consequently, the median radius of PNe increases from the previously known
sample, through MASH-II to MASH-I PNe, consistent with the population of  MASH PNe 
 generally representing objects that are more evolved (and hence 
more extended at a given distance) than those discovered  pre-MASH
(Parker et al.  2006).  This is in good agreement, for example, with the work
of Tajitsu et al.  (1999), whose photoionization models show increasing 
temperatures towards the lower left of the SMB plane.

\begin{figure}
\vspace{7cm}
\includegraphics{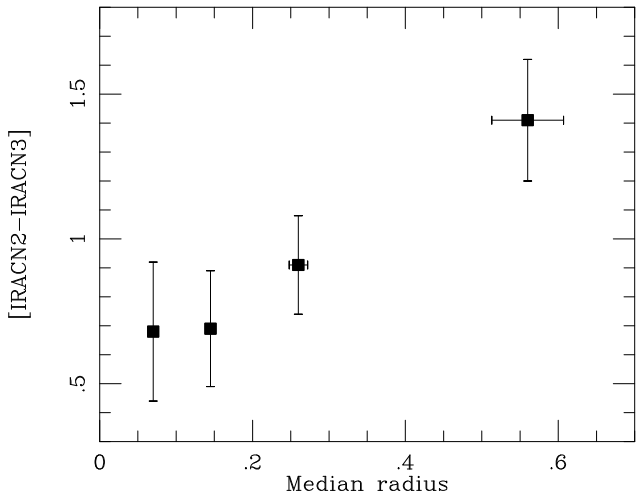}
\caption{The trend of IRAC [4.5$-$5.8] colour index with PN age using
median sample radius (in pc) as a proxy for each age bin. The index 
increases from very young to senile PNe.   \label{23age} } 
\end{figure}

\begin{figure}
\vspace{7cm}
\includegraphics{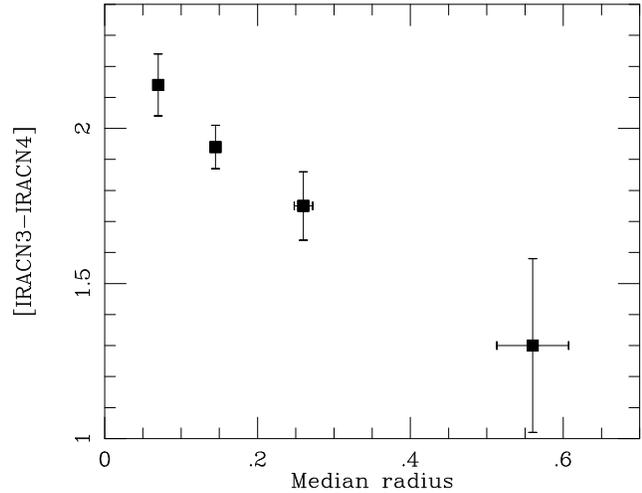}
\caption{The trend of IRAC [5.8$-$8.0] colour index with PN age using
median sample radius (in pc) as a proxy for each age bin.  This index decreases from 
very young to senile.  \label{34age} } 

\end{figure}

\subsection{Relationship between IRAC colours and nebular age}

Table~\ref{agecolours} investigates whether PNe also change their IRAC colours
as they age.  The sample of senile PNe is too small to provide meaningful 
median colours and was, therefore, merged with the old nebulae.  Fig.~\ref{23age} and
\ref{34age} show clear trends in both [4.5$-$5.8] and [5.8$-$8.0] colours as a function of PN age. 
Both plots can be understood in terms of an increasing contribution from the 
6.2-$\mu$m PAH band to the 5.8-$\mu$m flux as PNe and their PDRs expand with time. 
Recall (\S6.3) that MASH-II PNe have marginally smaller [4.5-5.8] than MASH-I PNe.
This now solidifies the impression that MASH-II PNe are truly younger and more 
compact than those found in MASH-I.   In Fig.~\ref{34age},
both relevant IRAC bands contain PAH emission suggesting that the relative 
brightnesses of these two bands change as PNe age.  Such an evolution implies
that the PN false colours vary with age.  The same interpretation applies equally
 to PNe in which the dominant emission mechanism is  by H$_2$ lines, 
and requires that the brightnesses of the strongest  lines such as those 
at 4.694, 5.510, and 6.907\,$\mu$m evolve accordingly.

{\bf Hot dusty haloes have been measured around some PNe (e.g.,Phillips 
\& Ramos-Larios (2005)), with dust temperatures in the range 700 to 1200K 
based on 2MASS colours. This could influence the IRAC photometry at 
3.6 microns. There is no supporting  evidence to date to show that 
stochastic heating of very tiny grains produces emission in the IRAC 
8.0 and 5.6 micron bands, which would require temperatures  between 
360K and 520K, respectively.}

\begin{table*}
\caption{IRAC colour indices for subsets of PNe according to age.
Median and standard error of the median (sem) are given together with the number 
of PNe that contribute to a colour and the total number of PNe available in 
that sample.  \label{agecolours} }
\begin{tabular}{cccccccc}
Subset&     $[3.6]-[4.5]$& $[3.6]-[5.8]$& $[3.6]-[8.0]$& $[4.5]-[5.8]$& $[4.5]-[8.0]$&   $[5.8]-[8.0]$&  Sample   \\
      &           No.&           No.&           No.&           No.&           No.&             No.&      No.\\    
\hline
Very Young& 0.84$\pm$0.13&  1.63$\pm$0.16&  3.87$\pm$0.17&  0.68$\pm$0.24&  2.96$\pm$0.24&  2.14$\pm$0.10& 18\\
    &          17&          18&          17&          17&          17&          17\\
Young& 0.90$\pm$0.11&    1.70$\pm$0.17& 3.59$\pm$0.16& 0.69$\pm$0.20& 2.44$\pm$0.19& 1.94$\pm$0.07&  44\\
     &     37&          39&          38&          38&          41&          39\\
Middle-aged& 0.64$\pm$0.18&  1.53$\pm$0.26&  3.52$\pm$0.27&  0.91$\pm$0.17&  2.69$\pm$0.16 &  1.75$\pm$0.11& 55\\
    &  32          31&          30&          33&          35&          33\\
Old+senile&     0.60$\pm$0.26& 1.81$\pm$0.26&  2.84$\pm$0.33& 1.41$\pm$0.21&  2.44$\pm$0.33&  1.30$\pm$0.28& 29\\
           16&           17&           16&          22&          23&         22\\
\hline
\end{tabular}   
\end{table*}

\section{The radio fluxes of the PN Sample}
Radio flux densities of the PNe were obtained from the MGPS-2 (Green 2002, Murphy et al.  2007)
and NVSS (Condon et al.  1998) surveys, together with data from Condon \& Kaplan (1998) and Luo, Condon \& Yin (1998).
MGPS-2 covers the sky south of $-$30$^\circ$ Dec. at 843\,MHz and NVSS
covers the sky north of $-$40$^\circ$ Dec. at 1.4\,GHz. The surveys
are comparable in resolution and sensitivity, and together they provide complete
coverage of the Galactic Plane at $\sim 1$ GHz. As in Paper-I, we treat
both surveys as if they offered data at a common frequency.
Both MGPS2 and NVSS provide online access to  point source catalogues.
For PNe that were not listed in these catalogues we nevertheless inspected the survey
images directly. If a credible, though weak, source, was apparent at the PN position, we
measured its flux using the MIRIAD task {\sc imfit}  to determine the
integrated emission above the  local background.

In Paper-I we concluded that PNe drawn from the sample consisting only of MASH nebulae were likely to be
optically thin near 1\,GHz.  Bojicic (2010) undertook a radio survey of PNe that are mostly    
MASH nebulae, previously undetected in NVSS and MGPS-2.  On the basis of the PN radio
brightness temperatures, these objects would also be expected to be optically thin  $\sim$1\,GHz
(Bojicic et al. (2010).   However, 30\,percent of the sample of PNe in the present paper
consists of compact, high-surface brightness PNe, known prior to MASH, many of them 
young.  One expects these PNe to be optically thick at low frequencies (e.g. Pazderska et al.  (2009; their fig.7).

Overall we found that 52\,percent (70/136) of our 136 PNe were detected around 1\,GHz.  As expected, the
detection rate is highest for those PNe discovered before MASH: 85\,percent (35/41).  For 
the generally more compact MASH-II nebulae the rate is 47\,percent (14/30) whereas 
for MASH-I it is 32\,percent (21/65), consistent with a
population of evolved PNe having larger diameters 
and weaker radio emission than previously known nebulae.

From our  optical spectra we have examined N$_e$ measured from the ratio of the red [S{\sc ii}] 
lines to test whether the highest density PNe are those with high MIR/radio simply 
by virtue of their radio optical depth $\sim$\,1GHz.  We found no correlation, 
perhaps because these lines cannot probe the highest density regions of PNe whose free-free 
emission dominates that from the more tenuous nebular regions.\\

\subsection{The MIR/radio flux ratio}
In Paper-I we interpreted differences found between the MIR/radio flux ratio for MASH-I PNe ($\sim$5)  and that for a set of 21 SEC 
PNe, (12), analyzed  by Cohen \& Green (2001)
as indicative of a progressive change in the ratio from previously known PNe to the more highly evolved MASH nebulae.  However, we have re-examined those 21
PNe and found that the MIR fluxes measured by MSX for a significant fraction of these objects were contaminated by other sources due to the location of the
nebulae in dense star fields at low latitudes.  These caused overestimates in the MIR/radio ratios.  This is rarely a problem for the measurement of bright sources, such as
H{\sc ii} regions, but accurate photometry with MSX of the faint sources that typify PNe requires caution in the Galactic plane. Therefore, we will not make use of the 
Cohen \& Green MSX fluxes for objects  affected by the crowding of the PN fields.  Thirteen of these 21 nebulae are  included in our new sample of
SEC PNe but now  all are  represented by IRAC images at high spatial resolution, eliminating the difficulties encountered by MSX photometry.
 The value for this ratio is now a more modest 8.4 for the SEC PNe.

The second column of Table~\ref{mirrad}  presents our analyses of this ratio for PNe drawn from the different discovery surveys, and by selecting PNe with
specific attributes such as the presence or absence of bipolarity in the nebular morphologies.  Bright PNe discovered prior to the  MASH and lying
within the GLIMPSE region contain a high fraction of bipolar nebulae.  A bipolar PN viewed close to its edge-on disk plane tends to be
brighter in the MIR than a non-bipolar object or a bipolar nebula viewed with its dusty disk face-on.   If a PN with 
a dust disk is seen face-on, the hot dust in the inner part of the disk closest to the star emits brightly in the NIR. 
For an edge-on {\bf optically thick}  disk the observer cannot see the inner disk directly. 
The radiation is is degraded to the IR and transferred through the disk to the cool outer edge, which is bright in the MIR because it subtends a larger angle than the hot dust 
and it peaks at longer wavelengths.

  One might interpret the large median MIR/radio value of 8.4 for the
33 SEC PNe as consistent with this picture,  but  the difference between this value and the MIR/radio ratio for all MASH nebulae (3.6) 
is statistically insignificant given the uncertainties.  The same situation prevails for the difference between all our bipolar and non-bipolar nebulae
(Table~\ref{mirrad}).  The highest value for MIR/radio (13) is found for the six PNe that lack any contrast relative to their ISM environment but this
sample is too small to provide a reliable median ratio.  
For the ensemble of IRAC colours, the MIR/radio flux ratio for our sample of PNe is also robustly represented by the median
value of 4.7$\pm$1.1 for the whole sample, virtually identical with the result derived in Paper-I for a much smaller group of PNe (4.6$\pm$1.2).
This ratio can also serve as a valuable and reliable new  
discriminant between PNe and H{\sc ii} regions. Diffuse   and compact H{\sc ii} regions present large MIR/radio ratios of $\sim$25 
while UCHIIs have even higher ratios with MIR/radio typically $\sim$42) (\S2.1.2) The significance of these differences between PNe and 
UCHIIs is 5\,$\sigma$, and between PNe and  compact and diffuse regions it is 4\,$\sigma$.

\subsection{The dependence of MIR/radio flux ratio on PN age}
It is of interest to probe whether changes in the MIR/radio ratio can be discerned as an effect of PN aging.  These ratios are found at the bottom
of the second column of Table~\ref{mirrad}.  There appear to be no significant differences between the ratios for very young, young, and middle-aged
PNe.  Only the sample of old PNe seems distinct, but only three nebulae contribute to the median ratio, and only two of 
these, PHR1437-5949 and PHR 1843-0232, have  high ratios of 20 and 21, respectively.  
Both are  PNe of large angular size, over 1\,arcmin in diameter, as expected for old nebulae.   
It is also possible that the 700\,mJy flux 
density of the latter, detected at 8.0\,$\mu$m in its direction, is contaminated by  ISM emission. 

That so few PNe are detected in the radio regime among these 29 old and senile objects suggests that 
most old PNe lose their radio emission faster than their PDRs are destroyed.  This is
consistent with their great faintness even in H$\alpha$.

\begin{table*}
\begin{center}
\caption{Median relationships between nebular excitation assessed as [N{\sc ii}]/H$\alpha$ and false IRAC colour for subsets of PNe
with different attributes.  For each quantity, median and standard error of the median (sem) are given. \label{mirrad}} 
\begin{tabular}{ccc}
Subset&     8.0$\mu$m/radio&  [N{\sc ii}]/H$\alpha$\\    
    &     PNe contributing/total & PNe contributing/total\\  
\hline
All PNe&    4.7$\pm$1.1&   2.1$\pm$0.3 \\
           &  66/136& 117/136  \\ 
MASH-I& 3.6$\pm$1.0&  2.4$\pm$0.4 \\       
       &     19/65&  56/65 \\
MASH-II& 3.9$\pm$0.8& 2.0$\pm$0.5 \\ 
      & 14/30& 29/30  \\
MASH& 3.6$\pm$0.9&  2.2$\pm$0.3\\ 
     &  33/95& 85/95    \\
Pre-mash&  8.4$\pm$2.8&  1.6$\pm$0.4\\
      &   33/41& 32/41    \\
\hline
Orange&  10$\pm$2.5&  2.3$\pm$0.5\\
      &      21/27& 24/27   \\
Red&  3.3$\pm$0.6& 2.1$\pm$0.3 \\
       &   30/38& 34/38 \\
Violet&  3.6$\pm$1.6& 0.7$\pm$0.2\\
      &   5/17& 15/17\\
O$+$R&  4.4$\pm$1.0& 2.2$\pm$0.3 \\  
     &    51/65& 58/65   \\
Contrast&  4.3$\pm$1.0&  2.1$\pm$0.8 \\ 
      &    56/78& 73/78  \\
No contrast&  13$\pm$3&  3.1$\pm$0.7\\
             &   6/31&  27/31 \\
\hline
Type~I&  10.5$\pm$2.4& 5.4$\pm0.7$   \\
      &          14/34  &    33/34\\
Non-Type~I&  4.3$\pm$1.0&   1.5$\pm$0.2\\  
      &           52/102&          84/102\\
\hline
BPNe&  7.4$\pm$2.4& 3.3$\pm$0.4 \\
      &   21/36&  34/34  \\
Non-BPNe&  3.6$\pm$1.0&  1.5$\pm$0.3 \\
       &  35/87&     74/87 \\
Elliptical& 3.6$\pm$1.0& 1.5$\pm$0.5\\
           &   29/65& 56/65  \\
Round&  4.0$\pm$0.8& 1.7$\pm$0.2  \\
          &   4/14&  11/14\\
\hline
Very young& 6.0$\pm$2&                    0.4$\pm$0.2\\
         &   16/18&      15/18\\
Young&     3.6$\pm$1.2 &     2.2$\pm$0.3\\
         &   29/44&  40/44\\
Middle-aged&  6$\pm$3&         2.5$\pm$0.5\\
         &   14/41&  37/41\\
Old&    20$\pm$5&     3$\pm$0.8\\
         &   3/25&  22/25\\ 
Old+senile&  &  3.0$\pm$1.5\\
         &     &    25/29 \\
\hline
\end{tabular} 
\end{center}  
\end{table*}


\section{False colour and excitation class}
The third column of Table~\ref{mirrad} explores the suggestion made in Paper-I that MIR false colours might serve as a proxy for optical
excitation of a PN, and hence also as a crude proxy for central star  temperature,
expressed as the ratio of [N{\sc ii}]/H$\alpha$ line strengths.  With our significantly enlarged sample of nebulae we 
see no progression of the median excitation from orange to red to violet.  However, the sole significant difference in this regard, between 
any pair of PN subgroups, is that between the combined orange+red sample and the violet false colour group. 
The formal difference is 4.2\,$\sigma$. Optical excitation class, represented by the ratio of [N{\sc ii}]/H$\alpha$, 
is related to  false colour in the sense that the combined orange and red PNe are distinct from violet PNe,  in the IRAC false-colour representation that we have adopted.

Although violet PNe constituted almost half the sample in Paper-I, they are now less than 13\,percent of our clean combined sample.
The dominance of the red and orange PNe suggests that these possibly represent the two most commonly 
found emission mechanisms in the IRAC wavelength range, namely PAH bands (red) or H$_2$ lines (orange).
The rarer violet PNe must represent a different population of nebulae. In Paper-I we suggested the possibility that violet nebulae might
represent two different PN types: objects dominated by H recombination lines, and nebulae of
high excitation where the interplay of emission lines in each band caused the violet appearance.
We had in mind the possibility of optically very low-excitation spectra for the former and 
very high excitation for the latter.  Note that some PNe in which only the H$\alpha$ line is seen in the red
without the common lines of [{N\sc ii}, [{S\sc ii}], are found to be
very high excitation PNe in which He lines are prominent.  It is, therefore, of interest that so few of the 
violet PNe have radio detections (less than one third: see Table~\ref{mirrad}, col(2)) 
perhaps reinforcing the possibility of a distinction based on the degree of ionization.  
There are 3 radio detections 
in the low-excitation group (low ratios of [{N\sc ii}]/H$\alpha$), and only one in the 
high-excitation group of violet PNe.  These numbers are so small that this cannot be regarded
as a trend.  The potential dichotomy of excitations suggested the importance of re-examining 
the violet objects.

The 17 violet PNe break down into three groups: 7 with [{N\sc ii}]/H$\alpha$ ratios $>$ 1;
8 with  [{N\sc ii}]/H$\alpha$ ratios $<$ 1; and two PNe for which we lack the requisite spectra.
The 7 high-excitation objects have a median ratio of 0.4$\pm$0.1, and the 8 low-excitation nebulae
a median ratio of 4.1$\pm$1.0.  Further, violet PNe showing only H$\alpha$ in emission are nebulae
of small diameter ($\leq$20\,arcsec), and tend to be elliptical.  Most of those with strong nitrogen lines  
tend to be bipolar. Apparently, false colour violet PNe come from two very different PN categories.
This is consistent with the unusually high error found in Table~\ref{colours} for the median
value of $[4.5]$-$[8.0]$ for false colour violet nebulae: 1.86$\pm$0.41.  This is the largest
error associated with any colour in this table and could also signify that two very different 
data sets are merged.
  
There is an additional clue from the Spitzer IRS  spectra of LMC PNe (Stanghellini et al.  2007).
Six PNe from the roughly 50 that have IRS spectra have 
IRAC false colour images that are violet.  All of these objects (SMP LMC numbers 10, 
28, 35, 45, 72, and 80) are characterized by ``featureless" MIR spectra, rising slowly to the red, 
and lacking any distinct emission peaks of either C-rich or O-rich mineralogy.  Despite this uniformity, 
these PNe span the range from intermediate to very high excitation MIR spectra. 
We have examined the optical spectra of these nebulae that were obtained  by 
Reid \& Parker (2006) and all 6 of these LMC PNe show only H lines.  [{N\sc ii}] 
is completely absent.  Many PNe have negligible [{N\sc ii}], [{O\sc ii }] and [{S\sc ii}]
lines.  These are very high excitation, optically thin PNe, and represent the most evolved
PNe in any sample (Richer et al.  2008). It is possible that at least some of these violet PNe are in fact
of such high excitation that one ought to acquire blue spectra and seek He{\sc ii}
lines to assess their excitation because the [{N\sc ii}]/H$\alpha$ ratio has definite
limitations. Reid \& Parker (2010) have revisited the three common methods for the excitation
class parameter for a large sample of LMC PNe.  They propose an improved scheme which
also uses the He{\sc ii} lines.  Most MASH PNe are too faint for meaningful blue spectra 
to be obtained and  we wanted to analyze all nebulae on a uniform basis, limiting the
current effort to the use of  [{N\sc ii}]/H$\alpha$.

By contrast, comparing the median excitation classes of Type~I and non-Type~I PNe indicates a substantial
difference (16\,$\sigma$) in the sense expected: Type~I nebulae have much higher excitation 
than non-Type~I nebulae, reflecting their more massive progenitors with the associated higher
nebular ionization and excitation.  
 
\section{Colour comparisons with the LMC PNe}   
We have compared the six IRAC median colours of our 136 PNe in GLIMPSE with
those of 588 LMC PNe.  We first examined the Leisy et al. (1997) LMC sample of PNe 
(IRAC data from Hora et al. (2008a), which are the LMC equivalents of the bright 
SEC sample of early discoveries in the Galaxy (our ``previously known" sample).  Then
we analyzed (Cohen et al. 2010, in prep.)  the combination of the Leisy PNe and
 the many fainter PNe discovered by Reid \& Parker (2006: RP) using IRAC 
measurements from the SAGE final data release (DR3, 2009-09-22 These source 
catalogues are based on mosaic photometry and are about 1 magnitude deeper 
than previous SAGE releases (Sewilo et al. 2009). Table~\ref{lmc}  compares our 
Galactic PN colours with those of the Leisy PNe and with the combined dataset of 
Leisy and RP PNe which best represents the LMC ensemble of nebulae.

Phillips \& Ramos-Larios (2009) tabulated IRAC data for a partial sampling of
the RP nebulae. Unfortunately they provide neither average nor median colours.
They state that LMC [4.5]-[5.8] indices are significantly higher than for Galactic PNe
while [5.8]-[8.0] indices are significantly lower. We disagree with both of 
these statements. We find no statistically meaningful difference between the
median IRAC colours of the 136 Galactic PNe and those of the sample of 
Leisy LMC PNe and only one meaningful difference between Galactic colours 
and those of the joint sample of all published LMC PNe.  In particular, the 
colour differences between median Galactic  and median LMC PNe in [4.5]-[5.8] 
and [5.8]-[8.0] are  only 0.6\,sigma and 2.9\,sigma, respectively. The sole significant 
difference occurs for [3.6]-[4.5], which is much smaller (5\,sigma away) for the Leisy+RP
PNe than in the Galaxy. However, this is attributable to contamination of the PN
photometry by Galactic and Magellanic field stars and even by the central stars of  
LMC PNe.  This phenomenon  was noted by Hora et al. (2008a) who refer to 
these as ``objects near zero colour". If we eliminate all LMC PNe whose 
[3.6]-[4.5] index is below 0.2,  the remaining 333  nebulae have  a median 
[3.6]-[4.5] of 0.76$\pm$0.02, indistinguishable  from the same colour indices 
for Galactic and Leisy PNe.  This reinforces the idea that the interlopers are field stars.
Despite the consequences for stellar evolution embodied in the very different metallicities of the Galactic
and LMC nebulae, no statistically meaningful colour differences are manifest.
It is not yet possible to probe more deeply into the corresponding MIR/radio
ratio for LMC PNe, because of the current paucity of radio continuum detections (Filipovic et al.  2009).

\begin{table*}
\caption{Comparison of  median IRAC colours (in Vega-based magnitudes) for all our PNe within GLIMPSE-I
with colours of LMC PNe from the Leisy et al.  (1997) tabulation, as measured in IRAC by Hora et al.  (2008a),
and 588 Leisy and RP PNe (IRAC from the SAGE project data release of 2009).
Median and standard error of the median (sem) are given together with the number of PNe that contribute 
to a colour.\label{lmc}}
\begin{tabular}{ccccccc}
&     $[3.6]-[4.5]$& $[3.6]-[5.8]$& $[3.6]-[8.0]$& $[4.5]-[5.8]$& $[4.5]-[8.0]$&   $[5.8]-[8.0]$ \\
Subset&                  No. &             No.&            No. &             No. &            No. & \\    
\hline
GLIMPSE-I&  0.81$\pm$0.08&  1.73$\pm$0.10&  3.70$\pm$0.11&  0.86$\pm$0.10&  2.56$\pm$0.11&  1.86$\pm$0.07\\   
        &    107&         110&         106&         115&         121&         116\\
Leisy&         0.76$\pm$0.03& 1.83$\pm$0.06&  3.39$\pm$0.07&  0.92$\pm$0.05&  2.62$\pm$0.07&  1.69$\pm$0.04\\
        &    174&         124&         141&         126&         145&         119\\
Leisy+RP&  0.58$\pm$0.02&  1.828$\pm$0.05&   3.327$\pm$0.08& 0.928$\pm$0.04&   2.464$\pm$0.06&  1.57$\pm$0.07\\
         &    433&        253&          254&         265&          266&         217\\
\hline
\end{tabular}
\end{table*}

\section{Conclusions}
 We have explored the contributions to a multi-wavelength analysis of
PNe offered by: the ratios of their red optical lines of H$\alpha$,  
[N{\sc ii}] and [S{\sc ii}]; electron densities derived from the 
[S{\sc ii}] doublet; IRAC false colours; excitation class, defined 
from the ratio of the [{N\sc ii}]/H$\alpha$ lines; direct overlays of
MIR image contours on H$\alpha$ images; and the MIR/radio ratio.
To place our goals in context with other efforts, we intend to pursue the possibility
of finding new PNe using Spitzer's IRAC wavelengths (3-8\,$\mu$m).  However,
both Wachter et al.  (2010) and Mizuno et al.  (2010) have recently 
described large numbers (over 200 and 362, respectively) of well-resolved 
bubbles found by visual inspection of Spitzer MIPS 24-$\mu$m images 
from the MIPSGAL  project (Carey et al.  2009).
Wachter et al. emphasize symmetric shells with defined peripheries and 
central sources.  These criteria lead to shells associated with Pop-I 
Wolf-Rayet stars and LBVs.  Mizuno et al. do not demand a central 
source and include filamentary, bipolar, and irregular objects.  Their
identifications are largely with PNe. Most of these bubbles are 
detected only at 24\,$\mu$m and have no extended IRAC counterparts.  
It will be of great interest to see whether our IRAC
approach will complement the 24-$\mu$m objects or overlap the MIPS 
detections.  But our new criteria will certainly help to select the best candidates for
NIR/MIR confirmatory spectroscopy.

In summary we are able to make the following points based on 
our investigations.
\begin{itemize}
\item  We have shown that 45\,percent of the ``known"  pre-MASH PNe in the 
GLIMPSE-I region  are likely contaminants. From repeated culls of 
MASH PNe we also estimate 
that our remaining degree of contamination by non-PNe is  5\,percent, almost an order 
of magnitude lower than that for heterogeneously catalogued PNe found prior 
to MASH in the same area. We believe that we have assembled a 
pure sample of 136 PNe to support these multi-wavelength studies that are, 
therefore, unaffected by any biases that these contaminants would introduce.

\item The total MIR extent of a PN should be treated as being at least
as large as the entire  region emitting in H$\alpha$  even if there is a 
bright MIR core close to the central star.  It is essential to measure a 
PN's entire MIR extent by going beyond the tracer of the ionized gas, 
whether thermal radio continuum or H$\alpha$, because the total MIR 
emission of a large PN, and its colours,  are quite different from the 
colours of the core.

\item The MIR/radio flux ratio and IRAC colour indices of Galactic 
PNe appear to be robust attributes, invariant among the many 
groups of PNe into which  we have divided our sample.   Somewhat 
suprisingly the colour indices of LMC PNe are statistically the 
same as those of Galactic PNe implying that any metallicity 
differences are not reflected in MIR PN attributes, or that these effects
contribute less than 0.10 mag to any index.

\item  Gatley's Rule appears to have no colour analogue in the MIR. 
We believe this is because the IRAC bands all contain H$_2$ lines.

\item We find 25 percent of our clean sample are Type I PNe, significantly
greater than a volume-limited estimate of 10 percent (Frew, Parker \&
Russeil 2006).  Therefore, GLIMPSE confirms that a Galactic
latitude-selected sample of PNe is dominated by higher-mass stellar
progenitors.

\item  Our comparisons show that PN MIR characteristics do not appear 
to vary with morphological subset, with few exceptions. PN IRAC colours 
cannot distinguish between bipolar and non-bipolar  nebulae. However, 
Type~I nebulae are strongly differentiated by their colours from non-Type~I PNe. 
PN optical excitation class does not correlate with IRAC colour  indices 
either but MIR false colour may be indicative, with a bimodal distribution 
of violet nebulae including both the lowest and the highest excitations.

\item  We have presented  IRAC colour-colour planes for PNe that are designed to
offer the largest metric distances between PNe and H{\sc ii} regions,
whether unresolved, diffuse or compact, and to minimize many other point source contaminants.

\item  We have  reduced by a factor of about three the diagnostic area in the SMB diagram in 
which bipolar PNe are located.

\item  We have assigned physical radii to all 136 PNe and have used these as 
proxies for PN age. On this basis we see a real trend between 
$[5.8]$-[$8.0]$ colour index and evolution which might itself serve as a 
more convenient practical determinant of evolutionary stage, at least capable of 
placing individual PNe in groups like ``young'', ``middle-aged'', etc.

\item  The median MIR/radio ratio for all PNe is 4.7$\pm$1.1 and 
it does not vary with PN evolutionary phase making PNe separable 
from their major contaminant, H{\sc ii} regions, whether compact,
diffuse, or ultra-compact.

\item From our enlarged sample we have determined that the MIR/radio ratio is 
not a proxy for PN excitation class, nor is PN false colour.

\item  We provide a table of 25 rejected objects in the GLIMPSE-I 
area that are not PNe in hopes that these will no longer confuse other researchers.

\item  The three clear IRAC false colour guises of PNe must indicate 
the mix of emission mechanisms that contribute to the IRAC appearance.
Regardless of the precise origin of these three colours, we argue that,
if a nebula displays any false colour other than these, this alone is
sufficient reason  to suspect that the nebula might not be a 
PN as such objects have invariably turned out to be contaminants 
when morphological, spectroscopic, NIR  and other diagnostic 
criteria are examined.

\item  PNe are known to suffer dimmings and brightenings and even to 
undergo outbursts (Shaw et al.  2007; Shaw, Rest \& Damke 2009).  Therefore, 
we prefer to use contemporaneous photometry to define the colours that
characterize PNe rather than to mingle IR data from several widely 
separated epochs.  

\item  We used the six IRAC colours from  Table~\ref{colours}, 
defined by the entire sample of PNE, to examine the same  median$\pm$3sems 
boxes as shown in our colour-colour planes.  We then sought point sources 
in the GLIMPSE-I catalog that satisfied the six constraints. This query recovered 
known PNe and returned a significant number of PN candidates.  These will
 be the subject of a future paper (Cohen et al. 2010, in prep.).
 
\end{itemize}

\section{Acknowledgments}
MC thanks NASA for supporting this work under ADP grants NNG04GD43G and NNX08AJ29G with
UC Berkeley.  MC is also grateful for support from the Distinguished
Visitor program at the Australia Telescope National Facility in Marsfield.
TM acknowledges the support of an ARC Australian Postdoctoral Fellowship (DP0665973).
The MOST is owned and operated by the University of Sydney, with support from the
Australian Research Council and Science Foundation within the School of Physics.
The National Radio Astronomy Observatory is a facility of the National Science Foundation 
operated under cooperative agreement by Associated Universities, Inc. 
This research made use of Montage, funded by the
National Aeronautics and Space Administration's Earth Science Technology
Office, Computational Technnologies Project, under Cooperative Agreement
Number NCC5-626 between NASA and the California Institute of Technology.
This research made use of SAOImage {\sc ds9}, developed by Smithsonian
Astrophysical Observatory.

\appendix
\section{Objects removed from our original sample}
Based on our careful re-evaluation of the provenance of our combined 
sample, especially important for the previously known pre-MASH PNe, 
we have been able to eliminate 25 objects from our compilation.
Some PNe were not observed at all by GLIMPSE due to the jagged 
boundaries in Galactic coordinates.  Nebulae without any IRAC 
observations are: PHR1654-4143, and MPA1525-5528.

See Table~\ref{rejects} for objects which we reject from our   
original sample of PN candidates. Note that no MASH-II objects 
were rejected as non-PNe in our surveyed area and only 5 MASH-I PNe 
(which tend to be more extended than MASH-II)  were removed 
while 20 ``known" PNe have beem eliminated. We suggest the most 
probable classification for each of these and document the 
reasons for such rejections.  Most are H{\sc ii} regions and 
can be recognized as such from their MIR morphology, false colours, 
high brightness, apparent size, large radio flux densities, and 
their characterization in the literature as bubbles. In the 
notes on the right of the table the abbreviations C06, K08, and 
P06 denote Churchwell et al.  (2006), Kwok et al.  2008, and 
Parker et al.  (2006). The first two  papers focus on bubbles, i.e., 
H{\sc ii} regions (C06); IRAC images of nebulae (K08). Note 
that P06 reject non-PNe typically because of morphology  and 
optical spectroscopy although other criteria such as 2MASS 
colours and MSX detections were also employed.

\begin{landscape} 
\begin{table*}

\caption{Objects rejected as non-PNe.\label{rejects}}
{\scriptsize
\begin{tabular}{llllllll}
Name&    RAJ2000&     DecJ2000&      GLON&      GLAT&  MASH& Nature&  Reason rejected/by whom\\ 
          &                &                  &          deg&  status&       deg&            &  \\
\hline
G295.7-00.2&       11 49 11.8&   $-$62 12 29&    295.747& $-$00.208& -&  H{\sc ii} region& MSX source; P06 morphology; K08,Fig.5\\
G296.31+00.68&  11 55 37.92&  $-$61 28 16.7& 296.319& 0.68& -& emission line star& PM 1-61 \\
G298.18-00.78&   12 09 01.2&  $-$63 15 58& 298.183& -0.786& -& compact H{\sc ii} region& He 2-77: this paper\\  
G298.4+00.6&      12 13 09.0&   $-$61 50 28&   298.431& +0.696& -&    H{\sc ii} region& this: morphology; very large radio flux; (K08, Fig.6 \\
G301.2+00.4&      12 37 09.6&   $-$62 23 10&   301.278& +0.440& -&    H{\sc ii} region& this paper: morphology;  K08, Fig.7 incorrectly 301.1+00.4 )\\
PHR1253-6350&   12 53  7.3&  $-$63 50 32  & 303.118&  $-$0.969& P&    Symbiotic star& Spectrum (Paper-I)\\
G306.4+ 00.2&     13 21 36.2&  $-$62 29 47&   306.407& +00.172& -&   ISM& due to diffuse ISM (this paper)   K08, Fig.8\\
G321.0-00.7&       13 22 05.5&   $-$62 24 06& 306.474& +00.259& -&   Reflection nebula?&  Fuzz around CPD-61 3629: K08, Fig.10\\ 
PHR1346-6116&   13 46 38.2&   $-$61 16 27&  309.516&  0.891&  L&    ISM&  This paper; see Fig.\,2\\ 
G309.80+00.56&  13 49 32.64&  $-$61 31 43.3& 309.8& 0.566& -& Pop-I WR nebula; WR 59, Simbad\\
G309.5-00.7&       13 49 52.1&  $-$62 49 41& 309.546& $-$0.709&   -&     H{\sc ii} region&  C06; very large radio flux; K08, Fig.9\\  
PHR1356-6139&   13 56 44.8& - $-$61 39 44 & 310.598&  0.234 & P&     ISM& due to diffuse ISM (this paper)\\
G321.3-00.3&       15 17 31.2&  $-$57 51 10& 321.392&  $-$0.311&  -&   H{\sc ii} region& PHR1517-5751rejected in Paper-I;  very large radio flux; C06; K08, Fig.11\\  
G321.20-00.82&   15 18 21.84&  $-$58 23 12.5& 321.204& $-$0.827& -& Post-AGB Star& PN PM 1-88: orange 2mass, MIR white; this paper;\\  
G328.5-00.5&       15 59 38.2&  $-$53 45 32& 328.573& $-$00.531& -& H{\sc ii} region& RCW99 P06 morphology;  very large radio flux; K08, Fig.12\\
PHR1603-5402&   16 03 41.4&  $-$54 02 04 & 328.841&   $-$1.131& L&    MSX 1arcmin source\\   
G329.6-00.4&       16 04 53&   $-$53 00.7&   329.654& $-$00.484& -& H{\sc ii} region& S69: rejected by P6 \& this paper;morphology;  very large radio flux;C06; K08, Fig.12\\
G332.5-00.1&       16 16 56.40&  $-$50 47 22.6&  332.528& $-$00.121& -&  H{\sc ii} region& PHR J1616-5047;P06 morphology; very large radio flux;K08, Fig.13\\
G331.72-01.01&   16 17 13.39&   $-$51 59 10.3& 331.727& $-$01.011& -&  Symbiotic star/B[e] & Mz~3: this paper \\ 
G333.7+00.3&      16 20 09.4&  $-$49 36 09&  333.726& +0.368& -&  H{\sc ii} region&  PHR J1620-4936 rejected (this paper) for morphology; very large radio flux;K08, Fig.13\\
G340.0+00.9&      16 43 16.00&   $-$44 35 18.0& 340.071& +0.927& -& H{\sc ii} region& YSO?  PHR J1643-4435; this paper; morphology; K08, Fig.14\\
PHR1644-4455&   16 44 36.2&   $-$44 55 23&  339.973&  0.528&  P& diffuse ISM& (this paper) \\
G018.6-00.0&       18 25 10.46&   $-$12 42 15.6& 18.656&  $-$0.058& -&   H{\sc ii} region& P06 for morphology; very large radio flux; K08, Fig.2\\
G026.47+00.02&   18 39 32.16&  $-$05 44 19.6& 26.470&  0.021& -& H{\sc ii} region& emission line MSX source; LBV candidate\\ 
G055.5-00.5&       19 36 11.0&   19 43 29&     55.492&   $-$0.495& -&   IRAS source& not a PN in SIMBAD\\
\hline
\end{tabular}
}
\end{table*}
\end{landscape}
\end{document}